\documentclass[12pt,journal,onecolumn,draftclsnofoot]{IEEEtran}
\newif\ifCLASSOPTIONromanappendices \CLASSOPTIONromanappendicestrue
\usepackage[utf8x]{inputenc}
\usepackage{a0size}         
\usepackage{amssymb}  
\usepackage{cuted}
\usepackage{flushend}
\usepackage{balance}
\usepackage{amsmath}
\usepackage{multicol}
\usepackage[english]{babel}    
\usepackage{epsfig} 
\usepackage{bm}
\usepackage{bbm}
\usepackage{amsfonts,color,amsthm,amsmath, lscape,graphics}
\usepackage{comment}
\usepackage{algorithm} 
\usepackage{algpseudocode}

\usepackage[font=footnotesize]{caption}
\usepackage[font=footnotesize]{subcaption}
\usepackage{subcaption}
\usepackage{epstopdf}
\usepackage{algorithm} 
\usepackage{algpseudocode}
\usepackage{cite}

\definecolor{awesome}{rgb}{1.0, 0.13, 0.32}

\addto\captionsenglish{}

\usepackage{graphicx}  
\theoremstyle{plain}

\usepackage{multirow} 
\usepackage{relsize}  
\DeclareMathAlphabet\mathbfcal{OMS}{cmsy}{b}{n}
\usepackage{bm}   
\usepackage{pifont}

\begin{document}
\title{Channel Estimation for Multicarrier Systems with Tightly-Coupled Broadband Arrays}
\author{Bamelak Tadele, Volodymyr Shyianov, Faouzi~Bellili, \IEEEmembership{Member, IEEE}, and Amine Mezghani, \IEEEmembership{Member, IEEE}
  \\\small Emails:  \{tadeleb, shyianov\}@myumanitoba.ca, \{Faouzi.Bellili, Amine.Mezghani\}@umanitoba.ca.
\thanks{The authors are with the Department of Electrical and Computer Engineering at the University of Manitoba, Winnipeg, MB, Canada.  This work was supported by the Discovery Grants Program of the Natural Sciences and Engineering Research Council of Canada (NSERC). Work accepted for publication in part in IEEE ICASSP'23 \cite{tadeleICASSP}.}}

\maketitle
\begin{abstract}
This paper develops a linear minimum mean-square error (LMMSE) channel estimator for single and multicarrier systems that takes advantage of the mutual coupling in antenna arrays. We model the mutual coupling through multiport networks and express the single-user multiple-input multiple-output (MIMO) communication channel in terms of the impedance and scattering parameters of the antenna arrays. We put forward a novel scattering description of the communication channel which requires only the scattering parameters of the arrays as well as the terminated far-field embedded antenna patterns. In multi-antenna single-carrier systems under frequency-flat channels, we show that neglecting the mutual coupling effects  leads to  inaccurate characterization of the channel and noise correlations. We also extend the analysis to frequency-selective multicarrier channels wherein  we further demonstrate that the coupling between the antenna elements within each array  increases the number of resolvable channel taps. Standard LMMSE estimators based on existing inaccurate channel models become sub-optimal when applied to the new physically consistent model. We hence develop a new LMMSE estimator that calibrates the coupling and optimally estimates the MIMO channel. It is shown that appropriately accounting for mutual coupling through the developed physically consistent model leads to remarkable performance improvements both in terms of channel estimation accuracy and achievable rate. We demonstrate those gains in a rich-scattering environment using a connected array of slot antennas both at the transmitter and receiver sides. 
\end{abstract}
\thispagestyle{empty}

\begin{IEEEkeywords}
Mutual Coupling, Wideband Channel Estimation, Multiport Communication, OFDM 
\end{IEEEkeywords}

\section{Introduction}
\subsection{Background and Motivation}
\IEEEPARstart{M}{ultiple}-Input Multiple-Output (MIMO) wireless systems, which use antenna arrays at the transmitter and receiver, leverage the spatial dimension of the channel to increase the data rate and/or improve the resilience to fading \cite{heath2018foundations}. Many of the theoretical promises brought by the MIMO technology are based on modeling antenna arrays with wide inter-element spacing (half-wavelength) wherein the electromagnetic effects of mutual coupling can be reasonably ignored. However, due to the ever-increasing demand for high data rates and reliability, future generations of wireless systems are evolving towards broadband massive MIMO technology for which many  traditional assumptions will be violated. Massive MIMO systems (integral in the evolution of 5G/6G communications \cite{saad2019vision}) where a large number of antennas are to be packed in a compact space, endure losses in fading diversity due to the excessive amount of mutual coupling between the antenna elements within the array. While tightly-coupled arrays have less spatial degrees of freedom, they enjoy larger bandwidth due to their ability to support both slowly and rapidly varying spatial current distributions, effectively creating an electrically connected structure \cite{cavallo2011connected,akrout2022super}. To that end, since future-generation wireless systems are expected to be super-wideband (i.e., with several octaves of operational bandwidth spanning both sub-6GHz and mmWave bands) the post-5G MIMO technology requires a drastic shift in the design of antenna arrays. The reason for the super-wideband requirement is that future antenna systems are expected to be multi-functional (i.e., used for both sensing and communication), multi-band, multi-standard, and multi-operator as opposed to the current technology \cite{saha2019multifunctional}. One promising antenna structure, to meet these demands, is the tightly-coupled connected array of slot antennas. Due to the fact that  it is physically connected, the slot antenna array is effectively a single aperture fed at multiple locations, which leads to tight coupling and an overall increase in the operational bandwidth of the antenna system \cite{cavallo2011connected}. Having a single antenna aperture fed at multiple locations is also more convenient from the perspective of analysis/design as well as implementation. As the number of antenna elements in the connected structure increases, enlarging the overall array aperture, the bandwidth keeps expanding with no theoretical low-frequency cut-off \cite{cavallo2011connected}. The practical fabrication of such arrays is simplified by the use of PCB technology. To that end, the analysis/design of the overall broadband MIMO system needs to employ channel models (both propagation and antenna) which are both tractable and consistent with the underlying physics.\\
Although antennas are fundamental devices for wireless transmissions, the analysis and design of MIMO systems have historically evolved around the basic precept of separating the mathematical abstractions of communication theory\footnote{Particularly the celebrated Shannon capacity formula for band-limited additive white Gaussian noise (AWGN) channels.}\cite{shannon1948mathematical} and the physical design considerations from antenna and electromagnetic theories \cite{balanis,jackson1999classical}. For instance, the wireless community assesses the performance of MIMO systems in terms of achievable rate and spectral efficiency criteria while the figure of merit for antenna design is the scattering parameters. Research effort has been recently made to bridge such assessment gap between communication and antenna communities, e.g., \textit{wave theory of information} \cite{franceschetti2017wave}, \textit{electromagnetic information theory} \cite{gruber2008new, migliore2008electromagnetics},  \textit{holographic MIMO} \cite{pizzo2020degrees}, and \textit{circuit theory for communication} \cite{ivrlavc2010toward}.
Multiport communication theory, first introduced in \cite{wallace2004mutual} and popularized by \cite{ivrlavc2010toward}, offers a consistent approach to incorporate the physics of radio-communication into the model of the channel matrix and the noise statistics. In this model, we have three interfaces between the transmitter and the receiver. The first and third interfaces consist of the multiport networks that aim to optimize (through different criteria) the link between the transmit/receive signals and their respective antennas. The middle interface is a multiport network that incorporates the physics of propagation as well as the coupling of the antennas in use. Together the communication channel is given by the relationship between the generator signal (voltage/current) and the load signal. This multiport model has led to new insights in beamforming \cite{ivrlavc2010toward}, was used to incorporate the impact of the antenna size on the achievable data rate \cite{shyianov2021achievable, akrout2022super} and was adopted to study the performance of near-field communication systems   \cite{akrout2021achievable}. Further, by merging multiport communication theory with information theory, the achievable rate criterion was used for the design of the matching networks in SISO systems \cite{taluja2010information,shyianov2021achievable} as well as the analysis of coupling in wideband SIMO systems \cite{saab2019capacity}.  
\subsection{Contribution}
The mutual coupling effects were previously investigated within the context of  carrier frequency offset estimation\cite{wu2010effects}  and the angle of arrival acquisition \cite{lui2010mutual}. Its impact on channel estimation was also explored in \cite{lu2007effect} where the authors incorporated the mutual coupling in a correlated channel model and have shown the performance degradation due to coupling. In this paper, we provide a more elaborate model that uncovers the effects of mutual coupling both in the channel and  noise statistics. As the network parameters of the antennas can be known a priori, we devise a scheme where one can leverage the mutual coupling to improve channel estimation compared to standard methods. The main contributions embodied by this paper are:
\begin{itemize}
    \item  a novel scattering description of the communication channel which requires only the scattering parameters of the arrays as well as the terminated far-field embedded antenna patterns. With the use of only terminated embedded antenna patterns, which can be easily measured as opposed to open/short circuit patterns, the novel description significantly simplifies the antenna design.
    \item a novel algorithmic solution to the MIMO channel estimation problem in single- and multi-carrier settings, which leverages the knowledge of the antenna scattering parameters to compensate for the effects of mutual coupling and array frequency response.
    \item in single-carrier systems under frequency-flat channels, we show an improvement of at least $10$ [\textrm{dB}] in normalized mean-squared estimation error with respect to standard LMMSE channel estimation. This substantial gain stems from incorporating the array mutual coupling in the channel estimation procedure. We also derive the achievable rate with the new channel estimation procedure and show that almost half of the achievable rate is lost if the mutual coupling is not taken into account.
    \item in a multi-carrier setting, the proposed algorithm also aims to equalize  the frequency selectivity of the antenna array. The gap between the standard LMMSE procedure and the novel antenna-aware procedure widens in presence of antenna array frequency selectivity. This is mainly due to the requirement to estimate a larger number of taps which increases the model complexity. Equivalently the remarkable enhancement in channel estimation performance  translates into appreciable achievable rate gains.
    \item in a multi-carrier setting, we also develop a joint space/frequency power allocation scheme and show that with the newly developed channel estimation procedure the power allocation is close to a perfect CSI scenario.
    
\end{itemize}
\subsection{Organization of the Paper and Notation}
We structure the rest of this paper as follows. In Section~\ref{sec:system model}, we introduce the model of the broadband MIMO wireless channel based on the impedance as well as scattering descriptions. In Section~\ref{sec:single carrier}, we introduce a single-carrier baseband equivalent channel model as well as develop an antenna-aware channel estimation procedure which we compare to the standard LMMSE channel estimation scheme. In Section \ref{sec:multi-carrier}, we present a baseband equivalent OFDM system and develop a multi-carrier antenna-aware estimation procedure. Finally, our simulation results are presented in Section~\ref{sec:simulation results}, where we describe the advantages of using the developed scattering description of the wireless channel as well as demonstrate performance advantages of the developed antenna-aware estimation procedure.
The following notation is used throughout this paper. Lower- and upper-case bold fonts (e.g., $\mathbf{x}$ and $\mathbf{X}$) are used to denote vectors and matrices, respectively, and vectors are in column-wise orientation by default. The $(m,n)$th entry of $\mathbf{X}$ is denoted as  ${X}_{m,n}$, and the $n$th element of $\mathbf{x}$ is denoted as $x_n$. Moreover,  $\{.\}^\textsf{T}$ and $\{.\}^\textsf{H}$ stand for the transpose and Hermitian (transpose conjugate) operators, respectively. The statistical expectation is denoted as $\mathbb{E}[\cdot]$ and the identity matrix is denoted as $\mathbf{I}$. Given any complex number,  $\Re\{\cdot\}$, returns its real part and we use $j$ to denote the imaginary unit (i.e., $j^{2}=-1$). Finally, $c$ denotes the speed of light in vacuum (i.e., $c \approx 3\times10^8\,[\textrm{m}/\textrm{s}]$), $T$ is the temperature in Kelvin, $\lambda$ is the wavelength, and $k_\text{b} = 1.38 \times 10^{-23}\, [\textrm{m}^{2}\, \textrm{kg} \,\textrm{s}^{-2}\, \textrm{K}^{-1}]$ is the Boltzmann constant.
\section{System Model}\label{sec:system model}
Multiport network analysis is a tool that we will utilize to characterize the properties of the antennas inside a circuit model. A generic multiport communication system consists of $N_t$ transmit generator voltages, described by $\boldsymbol{v}_{\textit{G}}$, that induce $N_r$ voltages at the receiver across the loads, which are described by $\boldsymbol{v}_{\textit{L}}$. This model can be written as:
\begin{equation}\label{eq:AA_MIMO_Channel}
    \boldsymbol{v}_{\textit{L}}(f) = \sqrt{\rho}{\mathbfcal{{H}_{\textit{eff}}}}(f)\boldsymbol{v}_{\textit{G}}(f) + \boldsymbol{n}(f), 
\end{equation}
where $\rho$ is the large-scale parameter and $\boldsymbol{n}(f)$ jointly represents the extrinsic and intrinsic noise sources at the receiver. We define ${\mathbfcal{{H}_{\textit{eff}}}}(f)$ as the ``effective channel" as it characterizes the antennas in use in addition to the propagation medium. We refer to the standard MIMO channel in the literature \cite{heath2018foundations} as the ``propagation channel" and denote it by $\mathbfcal{H}(f)$ which is determined based the embedded far-field patterns of the array elements under reference terminations as well as the propagation medium.
\begin{figure}[t]
    \centering
    \includegraphics[width=0.9\linewidth]{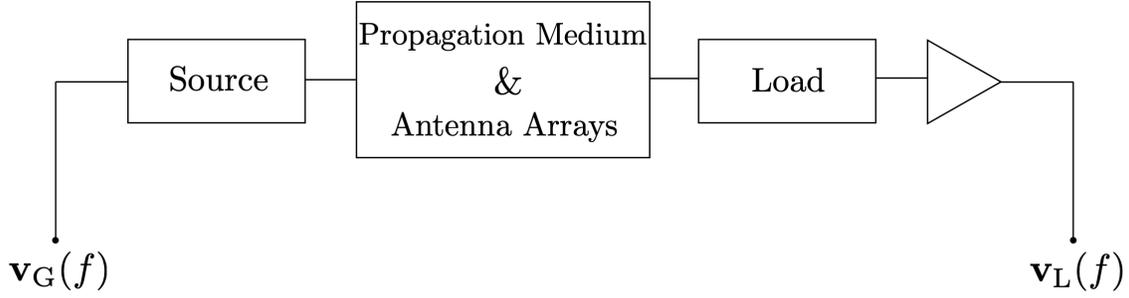}
    \vskip 0.1cm
    \caption{Generic Multi-port Communication System}  
    \label{fig:MIMO_Communication_System}
\end{figure}
\subsection{Impedance Description}
The first multiport network, seen as a "source" multiport, relates the transmit generator voltages, $\boldsymbol{v}_{\textit{G}}(f)$ to the voltages that will be induced on the $N_t$ transmit antennas, $\boldsymbol{v}_\textit{T}(f)$. This can be thought of as a feeding network for the transmit antennas and will incorporate the source impedances of the transmit voltages. This multiport can be represented by a single output impedance matrix $\mathbf{Z}_\textrm{S}(f)$. The middle multiport network represents the propagation medium and antenna arrays and is described through the impedance matrix, $\mathbf{Z}_\textrm{MIMO}(f)$ given by:
\begin{equation}\label{eq:mimo-V=ZI}
\Bigg[\begin{array}{l}
\boldsymbol{v}_{\textit{T}}(f) \\
\boldsymbol{v}_{\textit{R}}(f)
\end{array}\Bigg]~=~\underbrace{\Bigg[\begin{array}{cc}
\mathbf{Z}_{\text{T}}(f) & \mathbf{Z}_{\text{TR}}(f) \\
\mathbf{Z}_{\text{RT}}(f) & \mathbf{Z}_{\text{R}}(f)
\end{array}\Bigg]}_{\mathbf{Z}_{\text{MIMO}}(f)}\,\Bigg[\begin{array}{l}
\boldsymbol{i}_{\textit{T}}(f) \\
\boldsymbol{i}_{\textit{R}}(f)
\end{array}\Bigg].
\end{equation}
Here the transmit antenna impedance matrix is given by $\mathbf{Z}_{\text{T}}(f)$, the receive antenna impedance matrix is given by $\mathbf{Z}_{\text{R}}(f)$, and the propagation medium will be modeled through the use of $\mathbf{Z}_{\text{RT}}(f)$. By the unilateral approximation \cite{ivrlavc2010toward} we set $\mathbf{Z}_{\text{TR}}(f)=\mathbf{0}$. The interface between the receive antennas and the low-noise amplifiers (LNA) is represented by load matrix $\mathbf{Z}_\textrm{L}(f)$. This $\mathbf{Z}_\textrm{L}(f)$ is the input impedance looking at the load side and represents a ``load" multiport. We model the extrinsic noise collected at the receive antennas through the vector $\boldsymbol{v}_\mathrm{EN}(f)$ and the intrinsic amplifier noise is modeled through $\boldsymbol{v}_\mathrm{IN}(f)$. At last, we assume a single-source LNA with a gain of $\beta$. Finally, through basic circuit analysis, it is simple to show that
\begin{eqnarray}\label{eq:Impedance_MIMO_channel}
    \mathbfcal{H_{\textit{eff}}}(f) = \beta \mathbf{Z}_{\mathrm{L}}(f)[\mathbf{Z}_{\mathrm{R}}(f)+\mathbf{Z}_\mathrm{L}(f)]^{-1}
     \mathbf{Z}_{\mathrm{RT}}(f)[\mathbf{Z}_{\mathrm{T}}(f)+\mathbf{Z}_{\mathrm{S}}(f)]^{-1},
\end{eqnarray}
and 
\begin{equation}\label{Impedance_MIMO_Noise}
    \boldsymbol{n}(f)  = \boldsymbol{v}_\mathrm{IN}(f)+\beta\mathbf{Z}_{\mathrm{L}}(f)[\mathbf{Z}_{\mathrm{R}}(f)+\mathbf{Z}_{\mathrm{L}}(f)]^{-1}\boldsymbol{v}_\mathrm{EN}(f).
\end{equation}
We can write the mutual impedance, $\mathbf{Z}_{\mathrm{RT}}(f)$, that models the propagation channel as follows:
\begin{eqnarray}\label{eq:Z_RT-matrix}
    \mathbf{Z}_{\mathrm{RT}}(f) ~=~ \bigg[\textrm{diag}\left(\sqrt{\Re\{\mathbf{Z}_{\mathrm{R}}(f)\}}\right) 
    \mathbfcal{H}_{\rm OC}(f)\,\textrm{diag}\left(\sqrt{\Re\{\mathbf{Z}_{\mathrm{T}}(f)\}}\right)\bigg],
\end{eqnarray}
where $\mathbfcal{H}_{\rm OC}(f)$ is the propagation channel calculated based on open-circuit embedded pattern ($\neq \mathbfcal{H}(f)$). 
To compute the noise correlation matrix it is assumed that the noise voltage sources associated with different amplifiers are independent:
\begin{equation}
\mathbb{E}[\boldsymbol{v}_{\mathrm{IN}}(f)\,\boldsymbol{v}_{\mathrm{IN}}(f)^{\mathsf{H}}] = 4\beta^2\,k_b\,T\,\,(N_f - 1)R_{\textit{in}}\mathbf{I},
\end{equation}
where $(.)^{\mathsf{H}}$ denotes the conjugate-transpose of a matrix, $T$ is the noise temperature in Kelvin and $R_{\textit{in}}$ is the input impedance of the $N_r$ LNAs. 
The extrinsic noise correlation matrix, assuming the antenna array is in thermodynamic equilibrium with the environment is given by,
\begin{equation}
\mathbb{E}[\boldsymbol{v}_{\mathrm{EN}}(f)\boldsymbol{{v}}_{\mathrm{EN}}(f)^{\mathsf{H}}] = 4k_bT\mathbf{\Re\{\mathbf{Z}_{\textrm{R}}}(f)\}.
\end{equation}
With the above, the correlation of the total noise vector in (\ref{eq:AA_MIMO_Channel}), $\boldsymbol{R}_{\boldsymbol{n}}(f) \triangleq \mathbb{E}[\boldsymbol{n}(f)\boldsymbol{n}(f)^{\mathsf{H}}]$, can be determined as a sum of extrinsic and intrinsic noise correlations:
\begin{eqnarray}\label{eq:Impedance_noise_correlation}
     \boldsymbol{R}_{\boldsymbol{n}}(f) = \bigg[4k_bT\beta^{2}\mathbf{Z}_{L}(f)[\mathbf{Z}_{R}(f)+\mathbf{Z}_{L}(f)]^{-1}\Re\{\mathbf{Z}_{\textrm{R}}(f)\} &\nonumber \\ & \!\!\!\!\!\!\!\!\!\!\!\!\!\!\!\!\!\!\!\!\!\!\!\!\!\!\!\!\!\!\!\!\!\!\!\!\!\!\!\!\!\!\!\!\!\!\!\!\!\!\!\!\!\!\!\!\!\!\!\!\!\!\!\!\!\!\!\!\!\!\!\!\!\!\!\!\!\!\!\!\!\!\!\!\!\!\!\!\!\!\!\!\!\!\!\!\!\!\!\!\!\!\!\!\!\!\!\!\!\!\!\!\!\!\!\!\!\!\!\!\!\!\!\!\!\!\!\!\!\!\!
    \times [\mathbf{Z}_{R}(f)+\mathbf{Z}_{L}(f)]^{-\mathsf{H}}\mathbf{Z}_{L}(f)^{\mathsf{H}}\bigg] &\\ \nonumber& \!\!\!\!\!\!\!\!\!\!\!\!\!\!\!\!\!\!\!\!\!\!\!\!\!\!\!\!\!\!\!\!\!\!\!\!\!\!\!\!\!\!\!\!\!\!\!\!\!\!\!\!\!\!\!\!\!\!\!\!\!\!\!\!\!\!\!\!\!\!\!\!\!\!\!\!\!\!\!\!\!\!\!\!\!\!\!\!\!\!\!
    +~ 4k_bT\beta^2(N_f - 1)R_{in}\mathbf{I}.
\end{eqnarray}
\subsection{Scattering Description}
\noindent
In the scattering description, incident and reflected power waves at ports are related through scattering parameters. As the scattering parameters are measured using a load termination these are most useful for antenna analysis. The effective channel in the scattering description is given by \cite{jamaly2013multiport}:
\begin{eqnarray}\label{eq:Scattering_Channel_MIMO}
     \mathbfcal{H}_{\textit{eff}}(f)  = \frac{\beta}{4}\big[\mathbf{I}+\mathbf{S}_{\mathrm{L}}(f)\big]\big[\mathbf{I} - \mathbf{S}_{\mathrm{R}}(f)\mathbf{S}_{\mathrm{L}}(f)\big]^{-1}
    \mathbf{S}_{\mathrm{RT}}(f)\big[\mathbf{I}-\mathbf{S}_{\mathrm{S}}(f)\mathbf{S}_{\mathrm{T}}(f)\big]^{-1}\big[\mathbf{I}-\mathbf{S}_{\mathrm{S}}(f)\big],
\end{eqnarray}
where $\mathbf{S}_{\mathrm{RT}}(f)$ is given by:
\begin{eqnarray}\label{eq:Scattering_Z_rt}
    \mathbf{S}_{\mathrm{RT}}(f) = \sqrt{\mathbf{I} - \textrm{diag}(\mathbf{S}_{\mathrm{R}}(f)^{\mathsf{H}}\mathbf{S}_{\mathrm{R}}(f))}
    \mathbfcal{H}(f)\sqrt{\mathbf{I} - \textrm{diag}(\mathbf{S}_{\mathrm{T}}(f)^{\mathsf{H}}\mathbf{S}_{\mathrm{T}}(f))}.
\end{eqnarray}
Additionally, the corresponding noise correlation matrix is given by:
\begin{eqnarray}\label{eq:Scattering_noise_correlation}
    \boldsymbol{R}_{\boldsymbol{n}}(f) = \bigg[k_b T\beta^{2}Z_0[\mathbf{I}+\mathbf{S}_{\mathrm{L}}(f)][\mathbf{I} - \mathbf{S}_{\mathrm{R}}(f)\mathbf{S}_{\mathrm{L}}(f)]^{-1} &\\ \nonumber& \!\!\!\!\!\!\!\!\!\!\!\!\!\!\!\!\!\!\!\!\!\!\!\!\!\!\!\!\!\!\!\!\!\!\!\!\!\!\!\!\!\!\!\!\!\!\!\!\!\!\!\!\!\!\!\!\!\!\!\!\!\!\!\!\!\!\!\!\!\!\!\!\!\!\!\!\!\!\!\!\!\!\!\!\!\!\!\!\!\!\!\!\!\!\!\!\!\!\!\!\!\!\!\!\!\!\!\!\!
    \times [\mathbf{I}-\mathbf{S}_{\mathrm{R}}(f)^{\mathsf{H}} \mathbf{S}_{\mathrm{R}}(f)] [\mathbf{I} - \mathbf{S}_{\mathrm{R}}(f)\mathbf{S}_{\mathrm{L}}(f)]^{-\mathsf{H}} &\\ \nonumber& \!\!\!\!\!\!\!\!\!\!\!\!\!\!\!\!\!\!\!\!\!\!\!\!\!\!\!\!\!\!\!\!\!\!\!\!\!\!\!\!\!\!\!\!\!\!\!\!\!\!\!\!\!\!\!\!\!\!\!\!\!\!\!\!\!\!\!\!\!\!\!\!\!\!\!\!\!\!\!\!\!\!\!\!\!\!\!\!\!\!\!\!\!\!\!\!\!\!\! \times [\mathbf{I}+\mathbf{S}_{\mathrm{L}}(f)]^{\mathsf{H}}\bigg] + 4\beta^2k_\textrm{b}T(N_f - 1)R_{in}\mathbf{I}.
\end{eqnarray}
For the remainder of this paper we will exclusively use the scattering representation in (\ref{eq:Scattering_Channel_MIMO}) and (\ref{eq:Scattering_noise_correlation}) and assume that source and load terminations are designed such that $\mathbf{S}_{\mathrm{S}}$ = $\mathbf{0}$ and $\mathbf{S}_{\mathrm{L}}$ = $\mathbf{0}$. 
\section{Single-Carrier System}\label{sec:single carrier}
When the bandwidth $B$ is much smaller than the carrier frequency, $f_c$, and the channel is constant within the coherence block, the system model can be written in discrete time as:
\begin{equation}\label{eq: Discrete time model}
    \mathbf{y} ~=~ \sqrt{\rho}\,\mathbf{H}_{\textit{eff}} \, \mathbf{x} ~+~ \mathbf{n},
\end{equation}
where $\mathbf{y}$, $\mathbf{x}$, and $\mathbf{n}$ are discrete time samples of $\boldsymbol{v}_{\textit{L}}$, $\boldsymbol{v}_{\textit{G}}$ and $\boldsymbol{n}$ respectively. $\mathbf{H}_{\textit{eff}}$ is the discrete-time version of the channel in (\ref{eq:AA_MIMO_Channel}) at the carrier frequency\footnote{
For the single-carrier case, the channel and network parameters will be evaluated at frequency $f_c$ and we no longer write them as functions of frequency for compactness.}. In this case, $\mathbf{n}\sim\mathcal{CN}(\mathbf{{0}},\mathbf{R}_{\mathrm{n}})$ where $\mathbf{R}_{\mathrm{n}}$ is obtained from (\ref{eq:Scattering_noise_correlation}) as:
\begin{equation}\label{eq:Discrete_noise_correlation}
    \mathbf{R}_{\mathbf{n}} = B \boldsymbol{R}_{\boldsymbol{n}}(f_{c}).
\end{equation}
The channel, $\mathbf{H}_{\textit{eff}}$, can be decomposed as the product of three matrices as:
\begin{equation}\label{eq:discrete channel}
    \mathbf{H}_{\textit{eff}} ~\triangleq~ \mathbf{Q}\,\mathbf{H}\,\mathbf{F},
\end{equation}
where,
\begin{equation}\label{eq:F_andQ_definitions}
\begin{aligned}
    \mathbf{F} &~\triangleq~ \sqrt{\mathbf{I} - \textrm{diag}(\mathbf{S}_{\textrm{T}}^{\mathsf{H}}\mathbf{S}_{\textrm{T}})},\\
    \mathbf{Q} &~\triangleq~ \frac{\beta}{4}\sqrt{\mathbf{I} - \textrm{diag}(\mathbf{S}_{\textrm{R}}^{\textrm{H}}\mathbf{S}_{\textrm{R}})}.
\end{aligned}
\end{equation}
In a rich scattering environment the
propagation channel $\mathbf{H}$ can itself be written as a product of three matrices \cite{heath2018foundations}:
\begin{equation}\label{eq:AB_correlation_channel}
    \mathbf{H} = \mathbf{R_{\mathrm{R}}}^{1/2}\mathbf{H}_{\textrm{w}}\mathbf{R_{\mathrm{T}}}^{1/2},
\end{equation}
wherein the entries of $\mathbf{H}_{\textrm{w}}$ are modeled by i.i.d complex Gaussian random variables and
\begin{equation}\label{eq:receieve_and_transmit_correlations}
\begin{aligned}
    \mathbf{R}_{\mathrm{R}}^{1/2} &~\triangleq~ ({\mathbf{I} - \textrm{diag}(\mathbf{S}_{\textrm{R}}^{\mathsf{H}}\mathbf{S}_{\textrm{R}})})^{-1/2}(\mathbf{I} - \mathbf{S}_{\textrm{R}}^{\mathsf{H}}\mathbf{S}_{\textrm{R}})^{1/2},\\
    \mathbf{R}_{\mathrm{T}}^{1/2} &~\triangleq~ (\mathbf{I} - \mathbf{S}_{\textrm{T}}^{\mathsf{H}}\mathbf{S}_{\textrm{T}})^{1/2}({\mathbf{I} - \textrm{diag}(\mathbf{S}_{\textrm{T}}^{\mathsf{H}}\mathbf{S}_{\textrm{T}})})^{-1/2}.
\end{aligned}
\end{equation}
The transmit and receive spatial correlation matrices are obtained from the field identity which results from power conservation,
\begin{equation}
    (\mathbf{I} - \mathbf{S}^{\mathsf{H}}\mathbf{S})_{m,n}= \int_{4\pi}\mathbf{E}_m^{\textsf{H}}(\theta,\phi) \mathbf{E}_n(\theta,\phi) d\Omega
\end{equation}
where $\mathbf{E}(\theta,\phi)$ is the embedded far-field pattern of the array, either transmit or receive, with the corresponding phasing of the elements.
Further, it can be shown that the correlation matrix of the vectorized channel, $\mathrm{vec}(\mathbf{H})$, is given by:
\begin{equation}\label{eq:Kronecker_correlation}
    \mathbf{R}_{\mathbf{H}} = (\mathbf{R}_{\mathrm{T}}^{\mathsf{T/2}} {\mathbf{R}_{\mathrm{T}}^{\mathsf{1/2}}}^{*} )\otimes ({\mathbf{R}_{\mathrm{R}}}^{1/2} {\mathbf{R}_{\mathrm{R}}}^{\mathsf{H}/2}).
\end{equation}
Additionally, since the vectorized effective channel, $\mathrm{vec}(\mathbf{H}_{\textit{eff}})$, can be written as:
\begin{equation}\label{eq:SC_conversion_H_to_Heff}
    \mathrm{vec}(\mathbf{H}_{\textit{eff}})~=~ \mathbf{T}\,\mathrm{vec}(\mathbf{H}),
\end{equation}
where $\mathbf{T} ~\triangleq~ (\mathbf{F}^{\mathsf{T}} \otimes \mathbf{Q})$. We can write the correlation matrix of the vectorized effective channel as:
\begin{equation}\label{eq:Kronecker_correlation_effective}
    \mathbf{R}_{\mathbf{H}_{\textit{eff}}} = \mathbf{T}\,\mathbf{R}_{\mathbf{H}}\,\mathbf{T}^{\mathsf{H}}.
\end{equation}
%
%
\newline
\indent {To motivate the need for an antenna aware channel estimator, one that accounts for mutual coupling, we first demonstrate the limitations of the standard LMMSE antenna blind estimators that have traditionally ignored the coupling.} Assume that $N_p$ pilot signals, $\mathbf{x}(t)\in\mathbb{C}^{N_t}$, $t=1,\cdots,N_p$  are sent by the transmit array. The associated space-time matrix, $\mathbf{Y}\in\mathbb{C}^{N_r\times N_p}$ , of receive signals is given by: 
\begin{equation}\label{eq:classical_system_model}
    \mathbf{Y} ~=~ \sqrt{\rho}\, \mathbf{H}_{\textit{eff}}\, \mathbf{X} ~+~ \mathbf{N},
\end{equation}
where $\mathbf{X} = [\mathbf{x}(1),\ldots,\mathbf{x}(N_p)]$ and $\mathbf{N}=[\mathbf{n}(1),\ldots,\mathbf{n}(N_p)]\in\mathbb{C}^{N_r\times N_p}$ is the additive noise matrix. The noise vectors, $\mathbf{n}(t),~ t=1,\ldots,N_p$, are mutually independent and $\mathbf{n}(t)\sim\mathcal{{CN}}(\mathbf{0},\mathbf{R}_{\mathbf{n}}) ~\forall t$. By vectorizing (\ref{eq:classical_system_model}) it follows that: 
\begin{equation}\label{eq:classical_vec_system_model}
    \mathbf{\widetilde{y}} = \sqrt{\rho}\,\mathbf{A}\, \mathrm{vec}(\mathbf{H}_{\textit{eff}}) + \mathbf{\widetilde{n}},
\end{equation}
where $\mathbf{\widetilde{y}} \triangleq \mathrm{vec}(\mathbf{Y})$, $\mathbf{\widetilde{n}} \triangleq \mathrm{vec}(\mathbf{N})$, and $\mathbf{A}\triangleq (\mathbf{X}^{\mathsf{T}} \otimes \mathbf{I})$. From the model in (\ref{eq:classical_vec_system_model}) we can estimate $\mathrm{vec}(\mathbf{H}_{\textit{eff}})$ using the LMMSE estimator:
\begin{equation}
    \mathrm{vec}(\mathbf{\widehat{H}}_{\textit{eff}}) ~=~  \mathbf{W}_{\textrm{AB}}^{\mathsf{H}}\, \mathbf{\widetilde{y}},
\end{equation}
where:
\begin{equation}\label{eq:AB_estimator}
    \mathbf{W}_{\textrm{AB}} = \sqrt{\rho} \big(\widetilde{\mathbf{R}}_{\widetilde{\mathbf{n}}} + \rho \,\mathbf{A}\widetilde{\mathbf{R}}_{\mathbf{H_{\textit{eff}}}}\mathbf{A}^{\mathsf{H}} \big)^{-1}\mathbf{A}\widetilde{\mathbf{R}}_{\mathbf{H}_{\textit{eff}}}.
\end{equation}
While the model in (\ref{eq:classical_vec_system_model}) better characterizes the physics of communication, standard communication models ignore the channel and noise correlations stemming from mutual coupling of the transmit and receive antennas. This amounts to, wrongly, assuming $\widetilde{\mathbf{R}}_{\widetilde{\mathbf{n}}} = c_1 \mathbf{I}$ and $\widetilde{\mathbf{R}}_{\mathbf{H_{\textit{eff}}}} = c_2 \mathbf{I}$ for two constants $c_1$ and $c_2$. We will call the LMMSE estimator that does not take the mutual coupling into account the ``Antenna Blind" (AB) estimator and denote it by $\mathbf{W}_{\textrm{AB}}$. The constant $c_1$ is taken to be the noise power scaling factor in the perfectly matched case, i.e., the constant in (\ref{eq:Scattering_noise_correlation}) when $\mathbf{S}_{\mathrm{R}}$ = $\mathbf{0}$. The constant $c_2$ is the scale in the channel power given by the trace of the correlation matrix of $\mathrm{vec}(\mathbf{H}_{\textit{eff}})$ given by $\mathrm{tr}(\mathbf{R}_{\mathbf{H}_{\textit{eff}}})/(N_r N_t)$. These values are easily and accurately measured at the receiver.
\newline A useful metric to assess the performance of this estimator is the mean-squared error (MSE) matrix defined as:
\begin{equation}\label{eq:MSE_Matrix_defn}
    \mathbf{E}^{\textrm{AB}}_{\textit{eff}} ~\triangleq~ \mathbb{E}\{(\mathrm{vec}(\mathbf{H}_{\textit{eff}}) - \mathrm{vec}(\widehat{\mathbf{H}}_{\textit{eff}}))(\mathrm{vec}(\mathbf{H}_{\textit{eff}} - \mathrm{vec}(\widehat{\mathbf{H}}_{\textit{eff}}))^{\mathsf{H}} \}.
\end{equation}
This equals the covariance of the estimation error vector and fully describes the accuracy of the estimator. It can be shown that for the AB estimator,  this is given by:
\begin{eqnarray}\label{eq:SC_MSE_Matrix_effective}
    \mathbf{E}^{\textrm{AB}}_{\textit{eff}} ~=~ \mathbf{R}_{\mathbf{H}_{\textit{eff}}} ~-~ \rho \mathbf{R}_{\mathbf{H}_{\textit{eff}}}\, \mathbf{A}^{\mathsf{H}}\big(\widetilde{\mathbf{R}}_{\widetilde{\mathbf{n}}} + \rho \,\mathbf{A}\widetilde{\mathbf{R}}_{\mathbf{H_{\textit{eff}}}}\mathbf{A}^{\mathsf{H}} \big)^{-1}\mathbf{A}\widetilde{\mathbf{R}}_{\mathbf{H_{\textit{eff}}}}  &\\ \nonumber& \!\!\!\!\!\!\!\!\!\!\!\!\!\!\!\!\!\!\!\!\!\!\!\!\!\!\!\!\!\!\!\!\!\!\!\!\!\!\!\!\!\!\!\!\!\!\!\!\!\!\!\!\!\!\!\!\!\!\!\!\!\!\!\!\!\!\!\!\!\!\!\!\!\!\!\!\!\!\!\!\!\!\!\!\!\!\!\!\!\!\!\!\!\!\!\!\!\!\!\!\!\!\!\!\!\!\!\!\!\!\!\!\!\!\!\!\!\!\!\!\!\!\!\!\!\!\!\!\!\!\!\!\!\!\!\!\!\!\!\!\!\!\!\!\!\!\!\!\!\!
    ~-~ \rho \widetilde{\mathbf{R}}_{\mathbf{H_{\textit{eff}}}}^{\mathsf{H}}\, \mathbf{A}^{\mathsf{H}}\big(\widetilde{\mathbf{R}}_{\widetilde{\mathbf{n}}}^{\mathsf{H}} + \rho \,\mathbf{A}\widetilde{\mathbf{R}}_{\mathbf{H_{\textit{eff}}}}^{\mathsf{H}}\mathbf{A}^{\mathsf{H}} \big)^{-1}\mathbf{A}\mathbf{R}_{\mathbf{H}_{\textit{eff}}}^{\mathsf{H}} ~+~ \rho \widetilde{\mathbf{R}}_{\mathbf{H_{\textit{eff}}}}^{\mathsf{H}}\, \mathbf{A}^{\mathsf{H}}\big(\widetilde{\mathbf{R}}_{\widetilde{\mathbf{n}}}^{\mathsf{H}} + \rho \,\mathbf{A}\widetilde{\mathbf{R}}_{\mathbf{H_{\textit{eff}}}}^{\mathsf{H}}\mathbf{A}^{\mathsf{H}} \big)^{-1}
    &\\ \nonumber& \!\!\!\!\!\!\!\!\!\!\!\!\!\!\!\!\!\!\!\!\!\!\!\!\!\!\!\!\!\!\!\!\!\!\!\!\!\!\!\!\!\!\!\!\!\!\!
    *\big({\mathbf{R}}_{\widetilde{\mathbf{n}}} + \rho \,\mathbf{A}{\mathbf{R}}_{\mathbf{H_{\textit{eff}}}}\mathbf{A}^{\mathsf{H}} \big)\big(\widetilde{\mathbf{R}}_{\widetilde{\mathbf{n}}} + \rho \,\mathbf{A}\widetilde{\mathbf{R}}_{\mathbf{H_{\textit{eff}}}}\mathbf{A}^{\mathsf{H}} \big)^{-1}\mathbf{A}\widetilde{\mathbf{R}}_{\mathbf{H_{\textit{eff}}}} 
\end{eqnarray}
{Indeed the coupling mismatch effects are seen in the equation above as $\mathbf{R}_{\mathbf{H}_{\textit{eff}}} \neq  \widetilde{\mathbf{R}}_{\mathbf{H_{\textit{eff}}}}$ and ${\mathbf{R}}_{\widetilde{\mathbf{n}}} \neq \widetilde{\mathbf{R}}_{\widetilde{\mathbf{n}}}$. 
\newline
\indent It is important to decouple the effects of the antennas from the propagation medium and tailor estimation algorithms to estimate the truly unknown propagation channel.}
%
%
To this end, we start by rewriting (\ref{eq:classical_system_model}) in terms of the expansion in (\ref{eq:discrete channel}):
\begin{equation}\label{eq:AA_system_model}
    \mathbf{Y} ~=~ \sqrt{\rho}\,\mathbf{Q}\,\mathbf{H}\,\mathbf{F}\mathbf{X} ~+~ \mathbf{N}, 
\end{equation}
where $\mathbf{N}$ and $\mathbf{X}$ are the same as in (\ref{eq:classical_system_model}) and $\mathbf{H}$ is given in (\ref{eq:AB_correlation_channel}). At the receiver, we whiten the noise by using the Cholesky decomposition, $\mathbf{R}_{\mathbf{n}} = \mathbf{L}\mathbf{L}^{\mathsf{H}}$, by multiplying the received signal by $\mathbf{L}^{-1}$, that is:
\begin{equation}
    \mathbf{Y'} ~=~ \sqrt{\rho}\,(\mathbf{L}^{-1}\mathbf{Q})\,\mathbf{H}\,(\mathbf{FX}) ~+~ \mathbf{N'},
\end{equation}
where $\mathbf{Y'}\triangleq \mathbf{L}^{-1}\mathbf{Y}$ and $\mathbf{N'} \triangleq \mathbf{L}^{-1}\mathbf{N}$. Now the columns, $\{\mathbf{n}'(t)\}_{t=1}^{N_p}$, of $\mathbf{N'}$ are mutually independent and $\mathbf{n}'(t)\sim\mathcal{{CN}}(\mathbf{0},\mathbf{I}) ~\forall t$. We now apply the $\mathrm{vec}$(.) operator to obtain:
\begin{equation}\label{eq:AA_vec_system_model}
    \mathbf{\widetilde{y}'} = \sqrt{\rho} \mathbf{A'}\mathrm{vec}(\mathbf{H}) + \mathbf{\widetilde{n}'},
\end{equation}
where $\mathbf{\widetilde{y}'} \triangleq \mathrm{vec}(\mathbf{Y'})$, $\mathbf{\widetilde{n}'} \triangleq \mathrm{vec}(\mathbf{N'})$ and $$\mathbf{A'} ~\triangleq~ (\mathbf{FX})^{\mathsf{T}} \otimes (\mathbf{L}^{-1}\mathbf{Q}).$$ We can now use LMMSE estimation to reconstruct $\mathrm{vec}(\mathbf{H})$ from (\ref{eq:AA_vec_system_model}). By doing so we have incorporated the mutual coupling effects in estimation and will call this estimator the ``Antenna Aware" (AA) estimator given by:
\begin{equation}\label{eq:AA_estimator}
    \mathbf{W}_{\textrm{AA}} = \sqrt{\rho}(\mathbf{R_{\widetilde{n}'}} + \rho \mathbf{A'}\mathbf{R}_{\mathbf{H}}\mathbf{A'}^{\mathsf{H}})^{-1}\mathbf{A'}\mathbf{R}_{\mathbf{H}},
\end{equation}
where now we have $\mathbf{R_{\widetilde{n}'}} = \mathbf{I}$ and $\mathbf{R}_{\mathbf{H}}$ is given in (\ref{eq:Kronecker_correlation}). As the AA estimator estimates $\mathrm{vec}({\mathbf{H}})$, we first write the MSE matrix of $\mathrm{vec}(\mathbf{H})$ as:
\begin{equation}\label{eq:SC_MSE_H}
    \mathbf{E}^{\textrm{AA}} = \mathbf{R}_{\mathbf{H}} - \rho \mathbf{R}_{\mathbf{H}} \mathbf{A'}^{\mathsf{H}}(\mathbf{R_{\widetilde{n}'}} + \rho \mathbf{A'}\mathbf{R}_{\mathbf{H}}\mathbf{A'}^{\mathsf{H}})^{-1}\mathbf{A'}\mathbf{R}_{\mathbf{H}}.
\end{equation}
Finally, using the relation in (\ref{eq:SC_conversion_H_to_Heff}) we can write the MSE matrix of $\mathrm{vec}({\mathbf{H}_{\textit{eff}}})$ as:
\begin{equation}\label{eq:SC_MSE_Heff}
    \mathbf{E}^{\textrm{AA}}_{\textit{eff}} = \mathbf{T}\,\mathbf{E}^{\textrm{AA}}\,\mathbf{T}^{\mathsf{H}}.
\end{equation}

\section{Multicarrier Systems}\label{sec:multi-carrier}
In OFDM-based transmission, assuming the appropriate length of the cyclic prefix has been chosen, a given subcarrier will experience a frequency flat channel in the frequency domain. That is, the receiver observation $\bm{\mathsf{y}} \in \mathbb{C}^{N_r}$ of the symbol $\bm{\mathsf{x}} \in \mathbb{C}^{N_t}$ at subcarrier $k \in \{0,\hdots, K-1\}$ can be written as:
\begin{equation}\label{eq:OFDM_receieved signal}
    \bm{\mathsf{y}}[k] ~=~ \sqrt{\rho}\,\bm{\mathsf{H}}_{\textit{eff}}[k]\,\bm{\mathsf{x}}[k] ~+~ \bm{\mathsf{n}}[k],
\end{equation}
where $\bm{\mathsf{n}}[k]$ is the noise at the receiver. This relationship is also the same as taking a discrete-frequency sample of (\ref{eq:AA_MIMO_Channel}) where the $k^{th}$ subcarrier is sampled at frequency $f_{k}$. Similar to the single-carrier case, $\bm{\mathsf{n}}[k]\sim\mathcal{CN}(\mathbf{{0}},\bm{\mathsf{R}}_{\mathrm{n}}[k])$ where $\bm{\mathsf{R}}_{\mathrm{n}}[k]$ is obtained from (\ref{eq:Scattering_noise_correlation}) as:
\begin{equation}\label{eq:OFDM_Discrete_noise_correlation}
    \bm{\mathsf{R}}_{\mathrm{n}}[k] ~=~ (B/K) \boldsymbol{R}_{\boldsymbol{n}}(f_{k}).
\end{equation}
The channel at the $k^{th}$ subcarrier, can be written as:
\begin{equation}\label{eq:Effective_Channel_defn}
    \bm{\mathsf{H}}_{\textit{eff}}[k] ~\triangleq~ \bm{\mathsf{Q}}[k]\,\bm{\mathsf{H}}[k]\,\bm{\mathsf{F}}[k]
\end{equation}
where now\footnote{For compactness, we combine the terms from the correlation across the antennas and the part from the multiport network.}
\begin{equation}\label{eq:OFDM_F_andQ_definitions}
\begin{aligned}
    \bm{\mathsf{F}}[k] ~&~\triangleq~ \Big(\mathbf{I} - \mathbf{S}_{\textrm{T}}(f_{k})^{\mathsf{H}}\mathbf{S}_{\textrm{T}}(f_{k})\Big)^{-1},\\
    \bm{\mathsf{Q}}[k] ~&~\triangleq~ \frac{\beta}{4}\Big(\mathbf{I} - \mathbf{S}_{\textrm{R}}(f_{k})^{\mathsf{H}}\mathbf{S}_{\textrm{R}}(f_{k})\Big)^{-1}.
\end{aligned}
\end{equation}
Based on measurements of the PDP (performed independently of the antennas in use) an effective number of taps $L$ can be determined. These taps are due to the frequency selectivity of the propagation medium and as such, we can determine $\bm{\mathsf{H}}[k]$ from $L$ time-domain taps $\{\mathbf{H}[\ell]\}_{\ell = 0}^{L-1}$ using Discrete-Fourier Transform (DFT), i.e.,
\begin{equation}\label{eq:DFT_defintion}
    \bm{\mathsf{H}}[k]~=~\sum_{\ell=0}^{L-1}\mathbf{H}[\ell]\,e^{-j\,\frac{2\pi}{K}\ell k} ~~~~ k = 0,\hdots,K-1.
\end{equation}
As the distribution of the PDP will not alter our results significantly, for simplicity we assume a uniform PDP, where the entries of $\mathbf{H}[\ell]$ are i.i.d complex Gaussian random variables and the $\{\mathbf{H}[\ell]\}_{\ell = 0}^{L-1}$ are independent of one another. Indeed this is well justified in a rich-scattering model where a large number of paths contribute to each tap such that the power across delay should not vary drastically.
%
%
\\
\indent {As in the single-carrier case, we first demonstrate the limitation of using antenna blind estimators and thereafter derive an optimal antenna aware LMMSE estimator.} Channel estimation can be performed in the frequency domain by inserting a pilot sequence over the $K$ subcarriers and using OFDM-based transmission. Additionally, if we assume that the channel is time-invariant we can also use pilots at $L_t$ instances in time. That is, for $k \in \{0,\ldots, K-1\}$ and $t \in \{0,\ldots,L_t - 1\}$ the received signal at subcarrier $k$ and time $t$ can be written as:
\begin{equation}\label{eq:OFDM_classical_estimation_model}
    \bm{\mathsf{y}}[k,t] ~=~ \sqrt{\rho}\,\bm{\mathsf{H}}_{\textit{eff}}[k]\,\bm{\mathsf{x}}[k,t] ~+~ \bm{\mathsf{n}}[k,t].
\end{equation}
Note here that the noise is uncorrelated in frequency and time and correlated in space according to (\ref{eq:OFDM_Discrete_noise_correlation}). 
In the standard approach, the DFT relationship (\ref{eq:DFT_defintion}) would be applied on $\bm{\mathsf{H}}_{\textit{eff}}[k]$ using $L$ taps such that the system model can be expressed as:
\begin{align}
\bm{\mathsf{y}}[k,t] &~=~ \sqrt{\rho}\,\bm{\mathsf{H}}_{\textit{eff}}[k]\,\bm{\mathsf{x}}[k,t] ~+~ \bm{\mathsf{n}}[k,t]\label{eq:received-signal-blind}\\
 &~\approx~ \sqrt{\rho}\left(\sum_{\ell=0}^{L-1}\mathbf{H}_{\textit{eff}}[\ell]\,e^{-j\,\frac{2\pi}{K}\ell k}\right)\,\bm{\mathsf{x}}[k,t] ~+~ \bm{\mathsf{n}}[k,t].\label{eq:OFDM-received-signal-with-taps_approx}
\end{align}
If we had neglected the coupling in which $\bm{\mathsf{F}}[k]$ and $\bm{\mathsf{Q}}[k]$ would be identity matrices then there would be no approximation in going from (\ref{eq:received-signal-blind}) to (\ref{eq:OFDM-received-signal-with-taps_approx}). However, as seen in (\ref{eq:Effective_Channel_defn}) multiplying by $\bm{\mathsf{F}}[k]$ and $\bm{\mathsf{Q}}[k]$ in the frequency domain would cause spreading in time requiring a larger number of taps to accurately model the channel. The standard LMMSE estimator would take the relationship in (\ref{eq:OFDM-received-signal-with-taps_approx}) as exact and its estimator would be derived under this assumption. Simplifying further we define:
\begin{equation}\label{eq:Concatenated_Time_Channel}
    \overline{\mathbf{H}}_{\textit{eff}} ~\triangleq~ \big[\mathbf{H}_{\textit{eff}}[0] ~\cdots~ \mathbf{H}_{\textit{eff}}[L-1]\big],
\end{equation}
and
\begin{equation}
 \bm{\mathsf{u}}[k]~\triangleq~ \Big[1~e^{-j\frac{2\pi}{K}k}~\cdots~e^{-j\frac{2\pi}{K}k(L-1)}\Big]^{\mathsf{T}},
\end{equation}
(\ref{eq:OFDM-received-signal-with-taps_approx}) becomes:
\begin{equation}\label{eq:OFDM_received-signal-compact}
    \bm{\mathsf{y}}[k,t] ~\approx~ \sqrt{\rho}\,\overline{\mathbf{H}}_{\textit{eff}}\,\left(\bm{\mathsf{u}}[k]\,\otimes\, \bm{\mathsf{x}}[k,t]\right) ~+~ \bm{\mathsf{n}}[k,t].
\end{equation}
By applying $\mathrm{vec}(\cdot)$ to both sides of (\ref{eq:OFDM_received-signal-compact}) and using the identity $\mathrm{vec}(\mathbf{ABC}) = \left(\mathbf{C}^\top\,\otimes\,\mathbf{A}\right)\,\mathrm{vec}(\mathbf{B})$, we obtain:
\begin{equation}
    \bm{\mathsf{y}}[k,t] ~\approx~ \Big(\bm{\mathsf{u}}[k]^{\top}\,\otimes\, \bm{\mathsf{x}}[k,t]^\top\otimes\,\mathbf{I}\Big)\,\mathrm{vec}\Big(\overline{\mathbf{H}}_{\textit{eff}}\Big) ~+~ \bm{\mathsf{n}}[k,t].
\end{equation}
Stacking the observations $\bm{\mathsf{y}}[k,t]$ first over the $K$ subcarriers and then over the $L_t$ time-domain points we get:
\begin{equation}\label{eq:matrix-form-ofdm-AB}
    \overline{\bm{\mathsf{y}}} ~\approx~ \sqrt{\rho}\,{\mathbf{B}}\, \mathrm{vec}\Big(\overline{\mathbf{H}}_{\textit{eff}}\Big)~+~{\overline{\bm{\mathsf{n}}}},
\end{equation}
where 
\begin{align}
    \overline{\bm{\mathsf{y}}} & ~\triangleq~ \big[\bm{\mathsf{y}}[0,0],\hdots,\bm{\mathsf{y}}[K-1,L_t -1]\big]^{\mathsf{T}},\label{eq:stacking_OFDM_rec_sig}\\
    \overline{\bm{\mathsf{n}}} & ~\triangleq~ \big[\bm{\mathsf{n}}[0,0],\hdots,\bm{\mathsf{n}}[K-1,L_t -1]\big]^{\mathsf{T}}\label{eq:stacking_OFDM_noise},
\end{align}
and the matrix ${\mathbf{B}}$ is defined as:
\begin{equation}
    {\mathbf{B}} ~\triangleq~ \left[\begin{array}{c}
    \bm{\mathsf{u}}\left[0\right]^{\top} \otimes \bm{\mathsf{x}}\left[0,0\right]^{\top} \otimes \mathbf{I} \\
    \bm{\mathsf{u}}\left[1\right]^{\top} \otimes \bm{\mathsf{x}}\left[1,0\right]^{\top} \otimes \mathbf{I} \\
    \vdots \\
    \bm{\mathsf{u}}\left[K-1\right]^{\top} \otimes \bm{\mathsf{x}}\left[K-1,L_{t}-1\right]^{\top} \otimes \mathbf{I}
    \end{array}\right].
\end{equation}
The LMMSE channel estimate of (\ref{eq:matrix-form-ofdm-AB}) is given by:
\begin{equation}\label{eq:OFDM_AB_estimate}
\textrm{vec}\Big(\widehat{\overline{\mathbf{H}}_{\textit{eff}}}\Big) ~=~\mathbf{W}_{\textrm{AB}}^{\mathsf{H}}~ \bar{\bm{\mathsf{y}}},
\end{equation}
where
\begin{equation}\label{eq:OFDM_AB_estimator}
    \mathbf{W}_{\textrm{AB}} ~=~\sqrt{\rho}\,(\widetilde{\mathbf{R}}_{\overline{n}} + \rho {\mathbf{B}}\widetilde{\mathbf{R}}_{{\boldsymbol{H}_{\textit{eff}}}}{\mathbf{B}}^{\mathsf{H}})^{-1}{\mathbf{B}}\widetilde{\mathbf{R}}_{{\boldsymbol{H}_{\textit{eff}}}},
\end{equation}
where, similar to the single-carrier case, $\widetilde{\mathbf{R}}_{\overline{n}} = c_{3}\mathbf{I}$ and $\widetilde{\mathbf{R}}_{{\boldsymbol{H}_{\textit{eff}}}} = c_{4}\mathbf{I}$ are the (incorrectly mismatched) correlations of ${\overline{\bm{\mathsf{n}}}}$ and $\mathrm{vec}\Big(\overline{\mathbf{H}}_{\textit{eff}}\Big)$ respectively. The true noise correlation matrix ${\mathbf{R}}_{\overline{n}}$ can be found by first concatenating the noise correlation matrices (\ref{eq:OFDM_Discrete_noise_correlation}) of the $K$ subcarriers in a block diagonal matrix $\mathbf{R}_{\textrm{K}}$, i.e.,
\begin{equation}
    \mathbf{R}_{\textrm{K}} ~=~ \textrm{blockdiag}(\bm{\mathsf{R}}_{\mathrm{n}}[0],\bm{\mathsf{R}}_{\mathrm{n}}[1],\hdots,\bm{\mathsf{R}}_{\mathrm{n}}[K-1]).
\end{equation}
Since we use the $K$ subcarriers in $L_t$ instances of time we can express the total noise correlation matrix by repeating $\mathbf{R}_{\textrm{K}}$ in a block diagonal matrix:
\begin{equation}
    {\mathbf{R}}_{\overline{n}} ~=~ \textrm{blockdiag}\underbrace{(\mathbf{R}_{\textrm{K}},\mathbf{R}_{\textrm{K}},\hdots,\mathbf{R}_{\textrm{K}})}_{L_t \textrm{ repetitions}}.
\end{equation}
Again, similar to the single-carrier case we can take $c_{3}$ as the noise power scale factor with perfectly matched antennas at all subcarriers frequencies, i.e., the constant in (\ref{eq:Scattering_noise_correlation}) when $\mathbf{S}_{\mathrm{R}}(f_{k})$ = $\mathbf{0}$. The constant $c_4$ is chosen such that $\mathrm{tr}(\widetilde{\mathbf{R}}_{{\boldsymbol{H}_{\textit{eff}}}})$ is equal the channel power. It is most convenient to find the channel power of $\mathrm{vec}\Big(\overline{\mathbf{H}}_{\textit{eff}}\Big)$ in the frequency domain. We first start by defining:
\begin{equation}\label{eq:Concat_vec_definitions}
\begin{aligned}
    \overline{\bm{\mathsf{H}}}_{\textit{eff}} ~&~\triangleq~ \big[\bm{\mathsf{H}}_{\textit{eff}}[0] ~\cdots~ \bm{\mathsf{H}}_{\textit{eff}}[K-1]\big],\\
    \overline{\mathbf{H}}~&~\triangleq~ \big[{\mathbf{H}}[0] ~\cdots~ {\mathbf{H}}[L-1]\big],
\end{aligned}
\end{equation}
in which the vectorized versions are linearly related by:
\begin{equation}\label{eq:vec_fre_from_time}
    \mathrm{vec}\Big(\overline{\bm{\mathsf{H}}}_{\textit{eff}}\Big) ~=~ \mathbf{C}_{1}\,\mathrm{vec}\Big(\overline{\mathbf{H}}\Big),
\end{equation}
where
\begin{equation}
    \mathbf{C}_{1}~\triangleq~ \mathbf{D}\,\underbrace{\mathbf{P}_{\textrm{K}}\,\big( \mathbf{I}_{N_{t}} \otimes \big( \boldsymbol{\mathcal{F}}^{\mathsf{H}} \otimes \mathbf{I}_{N_{r}} \big)  \big)\, \mathbf{P}_{\textrm{L}}}_{\triangleq ~\mathbf{C}_{2}}.
\end{equation}
Here $\boldsymbol{\mathcal{F}}$ is a partial DFT matrix of size $L \times K$ with element $(m,n)$ given by:
$$\boldsymbol{\mathcal{F}}_{mn} ~\triangleq~ e^{j\frac{2\pi}{K}(m-1)(n-1)}.$$
$\mathbf{P}_{\textrm{L}}$ is a block diagonal permutation matrix that groups the taps next to each other to convert them to the frequency domain. $\mathbf{P}_{\textrm{K}}$ is another block diagonal permutation matrix that now groups the frequency components together. Finally $\mathbf{D}$ is used to get the effective channel in the frequency domain defined by:
\begin{equation}
    \mathbf{D}~\triangleq~ \textrm{blockdiag}\big( \bm{\mathsf{F}}^{\mathsf{T}}[0] \otimes \bm{\mathsf{Q}}[0]\,, \hdots, \bm{\mathsf{F}}^{\mathsf{T}}[K-1] \otimes \bm{\mathsf{Q}}[K-1] \big).
\end{equation}
As we assume a uniform PDP, the components of $\mathrm{vec}({\overline{\mathbf{H}}})$ are i.i.d Gaussian random variables with unit variance and in light of (\ref{eq:vec_fre_from_time}) we have:
\begin{equation}\label{eq:Covariance_channel_freq}
    \mathbf{R}_{\bm{\mathsf{H}}_{\textit{eff}}}~=~ \mathbf{C}_{1}\mathbf{C}_{1}^{\mathsf{H}}.
\end{equation}
Due to Parseval's theorem the power computed from (\ref{eq:Covariance_channel_freq}) is the same as the power in time and therefore $c_4 = \mathrm{tr}(\mathbf{R}_{\bm{\mathsf{H}}_{\textit{eff}}})/(N_r N_t L)$.
Using the Parseval theorem again we find the MSE matrix in the frequency domain based on $\mathrm{vec}\big(\overline{\bm{\mathsf{H}}}_{\textit{eff}}\big)$. To do this we write the true model in which $\overline{\bm{\mathsf{y}}}$ is generated by as:
\begin{equation}\label{eq:matrix-form-OFDM-ab_true}
        \overline{\bm{\mathsf{y}}} ~=~ \sqrt{\rho}\,{\mathbf{B'}}\, \mathrm{vec}\big(\overline{\bm{\mathsf{H}}}_{\textit{eff}}\big)~+~{\overline{\bm{\mathsf{n}}}},
\end{equation}
and the matrix ${\mathbf{B'}}$ is defined as:
\begin{equation}
    {\mathbf{B'}} ~\triangleq~ \left[\begin{array}{c}
    \bm{\mathsf{x}}\left[0,0\right]^{\top} \otimes \mathbf{I} \\
    \bm{\mathsf{x}}\left[1,0\right]^{\top} \otimes \mathbf{I} \\
    \vdots \\
    \bm{\mathsf{x}}\left[K-1,L_{t}-1\right]^{\top} \otimes \mathbf{I}
    \end{array}\right],
\end{equation}
stacking over the subcarriers first and then the time slots just like $\mathbf{B}$. Additionally to convert the time-domain estimate in (\ref{eq:OFDM_AB_estimate}) to the frequency domain we must multiply by $\mathbf{C}_{2}$. Indeed in this case we do not need to multiply by $\mathbf{D}$ as the AB estimator estimates the effective channel in time and already includes the antenna correlations. After some calculations, the MSE matrix is given by:
\begin{eqnarray}\label{eq:OFDM_AB_MSE}
    \mathbf{E}^{\textrm{AB}}_{\textit{eff}} ~=~ \mathbf{R}_{\bm{\mathsf{H}}_{\textit{eff}}} ~-~ \rho \mathbf{R}_{\bm{\mathsf{H}}_{\textit{eff}}}\, \mathbf{B'}^{\mathsf{H}}\big(\widetilde{\mathbf{R}}_{\widetilde{\mathbf{n}}} + \rho \,\mathbf{B}\widetilde{\mathbf{R}}_{\mathbf{H_{\textit{eff}}}}\mathbf{B}^{\mathsf{H}} \big)^{-1}\mathbf{B}\widetilde{\mathbf{R}}_{\mathbf{H_{\textit{eff}}}}\,\mathbf{C}_{2}^{\mathsf{H}}  &\\ \nonumber& \!\!\!\!\!\!\!\!\!\!\!\!\!\!\!\!\!\!\!\!\!\!\!\!\!\!\!\!\!\!\!\!\!\!\!\!\!\!\!\!\!\!\!\!\!\!\!\!\!\!\!\!\!\!\!\!\!\!\!\!\!\!\!\!\!\!\!\!\!\!\!\!\!\!\!\!\!\!\!\!\!\!\!\!\!\!\!\!\!\!\!\!\!\!\!\!\!\!\!\!\!\!\!\!\!\!\!\!\!\!\!\!\!\!\!\!\!\!\!\!\!\!\!\!\!\!\!\!\!\!\!\!\!\!\!\!\!\!\!\!\!\!\!\!\!\!\!\!\!\!
    ~-~ \rho \mathbf{C}_{2}\,\widetilde{\mathbf{R}}_{\mathbf{H_{\textit{eff}}}}^{\mathsf{H}}\, \mathbf{B}^{\mathsf{H}}\big(\widetilde{\mathbf{R}}_{\widetilde{\mathbf{n}}}^{\mathsf{H}} + \rho \,\mathbf{B}\widetilde{\mathbf{R}}_{\mathbf{H_{\textit{eff}}}}^{\mathsf{H}}\mathbf{B}^{\mathsf{H}} \big)^{-1}\mathbf{B'}\mathbf{R}_{\bm{\mathsf{H}}_{\textit{eff}}}^{\mathsf{H}} ~+~ \rho\, \mathbf{C}_{2}\,\widetilde{\mathbf{R}}_{\mathbf{H_{\textit{eff}}}}^{\mathsf{H}}\, \mathbf{B}^{\mathsf{H}}\big(\widetilde{\mathbf{R}}_{\widetilde{\mathbf{n}}}^{\mathsf{H}} + \rho \,\mathbf{B}\widetilde{\mathbf{R}}_{\mathbf{H_{\textit{eff}}}}^{\mathsf{H}}\mathbf{B}^{\mathsf{H}} \big)^{-1}
    &\\ \nonumber& \!\!\!\!\!\!\!\!\!\!\!\!\!\!\!\!\!\!\!\!\!\!\!\!\!\!\!\!\!\!\!\!\!\!\!\!\!\!\!\!\!\!\!\!\!\!\!\!\!\!\!\!\!\!\!\!\!\!\!\!\!\!\!\!\!\!\!\!\!\!\!\!\!\!\!\!\!
    *\big({\mathbf{R}}_{\widetilde{\mathbf{n}}} + \rho \,\mathbf{B'}\mathbf{R}_{\bm{\mathsf{H}}_{\textit{eff}}}\mathbf{B'}^{\mathsf{H}} \big)\big(\widetilde{\mathbf{R}}_{\widetilde{\mathbf{n}}} + \rho \,\mathbf{B}\widetilde{\mathbf{R}}_{\mathbf{H_{\textit{eff}}}}\mathbf{B}^{\mathsf{H}} \big)^{-1}\mathbf{B}\widetilde{\mathbf{R}}_{\mathbf{H_{\textit{eff}}}}\,\mathbf{C}_{2}^{\mathsf{H}}. 
\end{eqnarray}
%
{As in the single-carrier case, the correlation mismatch effects are seen above as $\mathbf{R}_{\bm{\mathsf{H}}_{\textit{eff}}} \neq  \widetilde{\mathbf{R}}_{\mathbf{H_{\textit{eff}}}}$ and ${\mathbf{R}}_{\widetilde{\mathbf{n}}} \neq \widetilde{\mathbf{R}}_{\widetilde{\mathbf{n}}}$. In multicarrier transmission, we now also have mismatch effects due incorrectly truncating the channel in time to $L$ taps, manifested in (\ref{eq:OFDM_AB_MSE}) as $\mathbf{B} \neq \mathbf{B'}$. 
\\
\indent Again, it is important to decouple the effects of the antennas from the effective channel and tailor estimation algorithms to estimate the propagation channel alone.}
To extend the antenna aware estimator for the OFDM system we start by writing the OFDM system with the decomposed channel, i.e.,
\begin{equation}\label{eq:OFDM_AA_System}
    \bm{\mathsf{y}}[k,t] ~=~ \sqrt{\rho}\,\bm{\mathsf{Q}}[k]\,\bm{\mathsf{H}}[k]\,\bm{\mathsf{F}}[k]\,\bm{\mathsf{x}}[k,t] + \bm{\mathsf{n}}[k,t],
\end{equation}
 Additionally, the covariance matrix for subcarrier $k$ can be decomposed in its cholesky decomposition, $\bm{\mathsf{R}}_{\mathrm{n}}[k] \triangleq \bm{\mathsf{L}}[k]\bm{\mathsf{L}}[k]^{\mathsf{H}}$ wherein the cholesky factors are now a function of $k$. Applying $\bm{\mathsf{L}}^{-1}[k]$ to the received signal $\bm{\mathsf{y}}[k,t]$ we get:
\begin{align}
    \bm{\mathsf{y}}'[k,t] &~\triangleq ~\bm{\mathsf{L}}^{-1}[k]\,\bm{\mathsf{y}}[k,t]\\
    &~=~ \sqrt{\rho}\,\bm{\mathsf{L}}^{-1}[k]\,\bm{\mathsf{Q}}[k]\,\bm{\mathsf{H}}[k]\,\bm{\mathsf{F}}[k]\,\bm{\mathsf{x}}[k,t] ~+~ \bm{\mathsf{n}}'[k,t],
\end{align}
where now $\bm{\mathsf{n}}'[k,t]\sim\mathcal{CN}(\mathbf{{0}},\mathbf{I})$. By defining $\bm{\mathsf{P}}[k]\triangleq\bm{\mathsf{L}}^{-1}[k]\bm{\mathsf{Q}}[k]$, and following similar steps as (\ref{eq:received-signal-blind})--(\ref{eq:matrix-form-ofdm-AB}) we get:
\begin{equation}\label{eq:matrix-form-ofdm-antenna-aware}
    \overline{\bm{\mathsf{y}}}^{\prime} ~=~ \sqrt{\rho}\,\mathbf{M}\, \textrm{vec}\Big(\overline{\mathbf{H}}\Big)~+~{\overline{\bm{\mathsf{n}}}'}.
\end{equation}
Here $\overline{\bm{\mathsf{y}}}^{\prime} $ and ${\overline{\bm{\mathsf{n}}}'}$ are defined similar to (\ref{eq:stacking_OFDM_rec_sig}) and (\ref{eq:stacking_OFDM_noise}) by stacking $\bm{\mathsf{y}}'[k,t]$ and $\bm{\mathsf{n}}'[k,t]$ first over the subcarriers and then over time slots. $\textrm{vec}\Big(\overline{\mathbf{H}}\Big)$ is the vectorized channel in the time domain, similar to the channel in (\ref{eq:matrix-form-ofdm-AB}), but now without the $\bm{\mathsf{Q}}[k]$ and $\bm{\mathsf{F}}[k]$ factors. The matrix $\mathbf{M}$ is defined as:
\begin{equation}\label{eq:OFDM_AA_sensing_matrix}
    \mathbf{M} ~\triangleq~ \left[\begin{array}{c}
    \bm{\mathsf{u}}\left[0\right]^{\mathsf{T}} \otimes (\bm{\mathsf{F}}[0]\,\bm{\mathsf{x}}\left[0,0\right])^{\mathsf{T}} \otimes \bm{\mathsf{P}}[0] \\
    \bm{\mathsf{u}}\left[1\right]^{\mathsf{T}} \otimes (\bm{\mathsf{F}}[1]\,\bm{\mathsf{x}}\left[1,0\right])^{\mathsf{T}} \otimes \bm{\mathsf{P}}[1] \\
    \vdots \\
    \bm{\mathsf{u}}\left[K-1\right]^{\mathsf{T}} \otimes (\bm{\mathsf{F}}[K-1]\,\bm{\mathsf{x}}\left[K-1,L_t -1\right])^{\mathsf{T}} \otimes \bm{\mathsf{P}}[K-1]
    \end{array}\right].
\end{equation}
\noindent Finally, we obtain the OFDM version of the LMMSE Antenna Aware estimator:
\begin{equation}\label{eq:OFDM_AA_estimate}
\textrm{vec}\left(\widehat{\overline{{\mathbf{H}}}}\right)=\mathbf{W}_{\mathrm{AA}}^{\mathsf{H}}~ \overline{\mathbf{y}}^{\prime},
\end{equation}
where
\begin{equation}\label{eq:OFDM_AA_estimator}
    \mathbf{W}_{\textrm{AA}} ~=~ \sqrt{\rho}\,(\bm{\mathsf{R}}_{\overline{\bm{\mathsf{n}}}'} ~+~ \rho\, \mathbf{M}\, \mathbf{R}_{{\overline{\mathbf{H}}}}\,\mathbf{M}^{\mathsf{H}})^{-1}\,\mathbf{M}\,\mathbf{R}_{{\overline{\mathbf{H}}}},  
\end{equation}
where $\mathbf{R}_{\overline{\mathbf{H}}}$ and $\bm{\mathsf{R}}_{\overline{\bm{\mathsf{n}}}'}$ denote the correlation matrices of $\mathrm{vec}\Big(\overline{\mathbf{H}}\Big)$ and $\overline{\bm{\mathsf{n}}}'$ respectively. Due to the noise whitening process, $\bm{\mathsf{R}}_{\overline{\bm{\mathsf{n}}}'} = \mathbf{I}$ and since channels are i.i.d $\mathbf{R}_{\overline{\mathbf{H}}} = \mathbf{I}$. Similar to the single-carrier case, we first write the MSE matrix of $\mathrm{vec}\Big(\overline{\mathbf{H}}\Big)$ as:
\begin{equation}\label{eq:OFDM_MSE_H}
    \mathbf{E}^{\textrm{AA}} ~=~ \mathbf{R}_{\overline{\mathbf{H}}} ~-~ \rho \,\mathbf{R}_{\overline{\mathbf{H}}} \mathbf{M}^{\mathsf{H}}(\bm{\mathsf{R}}_{\overline{\bm{\mathsf{n}}}'} ~+~ \rho \mathbf{M}\,\mathbf{R}_{\overline{\mathbf{H}}}\,\mathbf{M}^{\mathsf{H}})^{-1}\mathbf{M}\,\mathbf{R}_{\overline{\mathbf{H}}}.
\end{equation}
Using the relation in (\ref{eq:vec_fre_from_time}) we can write the MSE matrix of $\mathrm{vec}\big(\overline{\bm{\mathsf{H}}}_{\textit{eff}}\big)$ as:
\begin{equation}\label{eq:OFDM_MSE_AA_eff}
    \mathbf{E}^{\textrm{AA}}_{\textit{eff}} ~=~ \mathbf{C}_{1}\,\mathbf{E}^{\textrm{AA}}\,\mathbf{C}_{1}^{\mathsf{H}}.
\end{equation}


\section{Simulation Results}\label{sec:simulation results}
\subsection{Scattering/Impedance channel equivalence}
\begin{figure}[h!]
    \centering
    \includegraphics[width=0.7\linewidth]{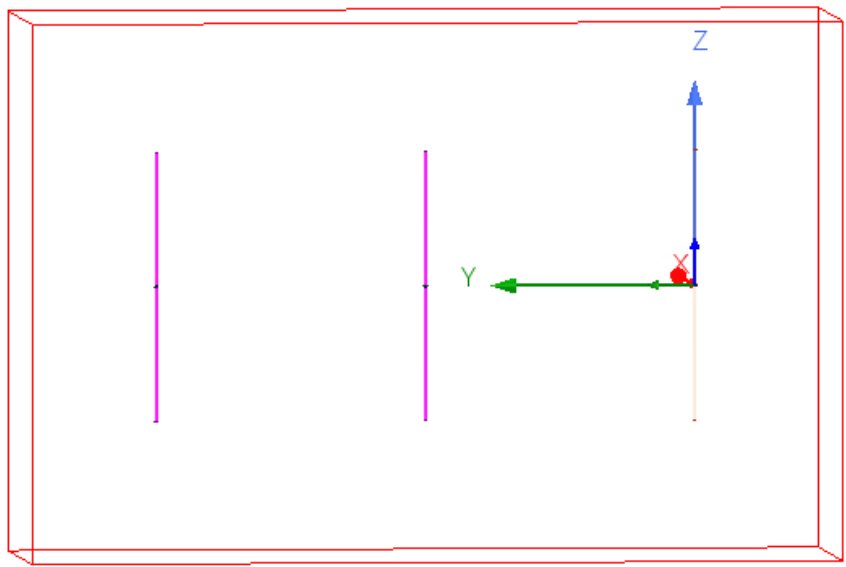}
    \vskip 0.1cm
    \caption{Array of two dipole antennas} 
    \label{fig: Dipole array}
\end{figure}

In this section, we will numerically demonstrate the equivalence between the MIMO channel description in (\ref{eq:Impedance_MIMO_channel}) using impedance parameters and the description in (\ref{eq:Scattering_Channel_MIMO}) using scattering parameters. To our best knowledge, the simple circuit-based channel model using scattering description in (\ref{eq:Scattering_Channel_MIMO}) did not yet appear in the literature. Most works on wireless network modeling with S-parameters either use a hybrid Impedance/Scattering description \cite{wallace2004mutual}, or utilize a field based approach with infinite expansion for the radiation pattern \cite{gately1968network}. The scattering description in (\ref{eq:Scattering_Channel_MIMO}) is simple since only finitely many basis functions are needed to represent the field, yet it is not hybrid and does not use the open/short circuit patterns which are much more difficult to measure in practice. To show the equivalence we simulate an array of two half-wavelength dipole antennas at $1$[\textrm{GHz}] both for the transmitter and the receiver. The antennas are assumed to have parallel configuration with half-wavelength spacing. The simple HFSS set-up is shown in Fig.\ref{fig: Dipole array}.
The considered propagation channel is Line-of-sight (LoS) link with the Friis pathloss which is given by,
\begin{equation}
    \mathbfcal{H}_{\rm OC/Term}(f) = \frac{c}{4\pi fd}G_{\rm OC/Term}\mathbf{1}\mathbf{1}^{\textsf{T}},
\end{equation}
\begin{figure}
\centering
\begin{subfigure}{.5\textwidth}
  \centering
  \includegraphics[width=1\linewidth]{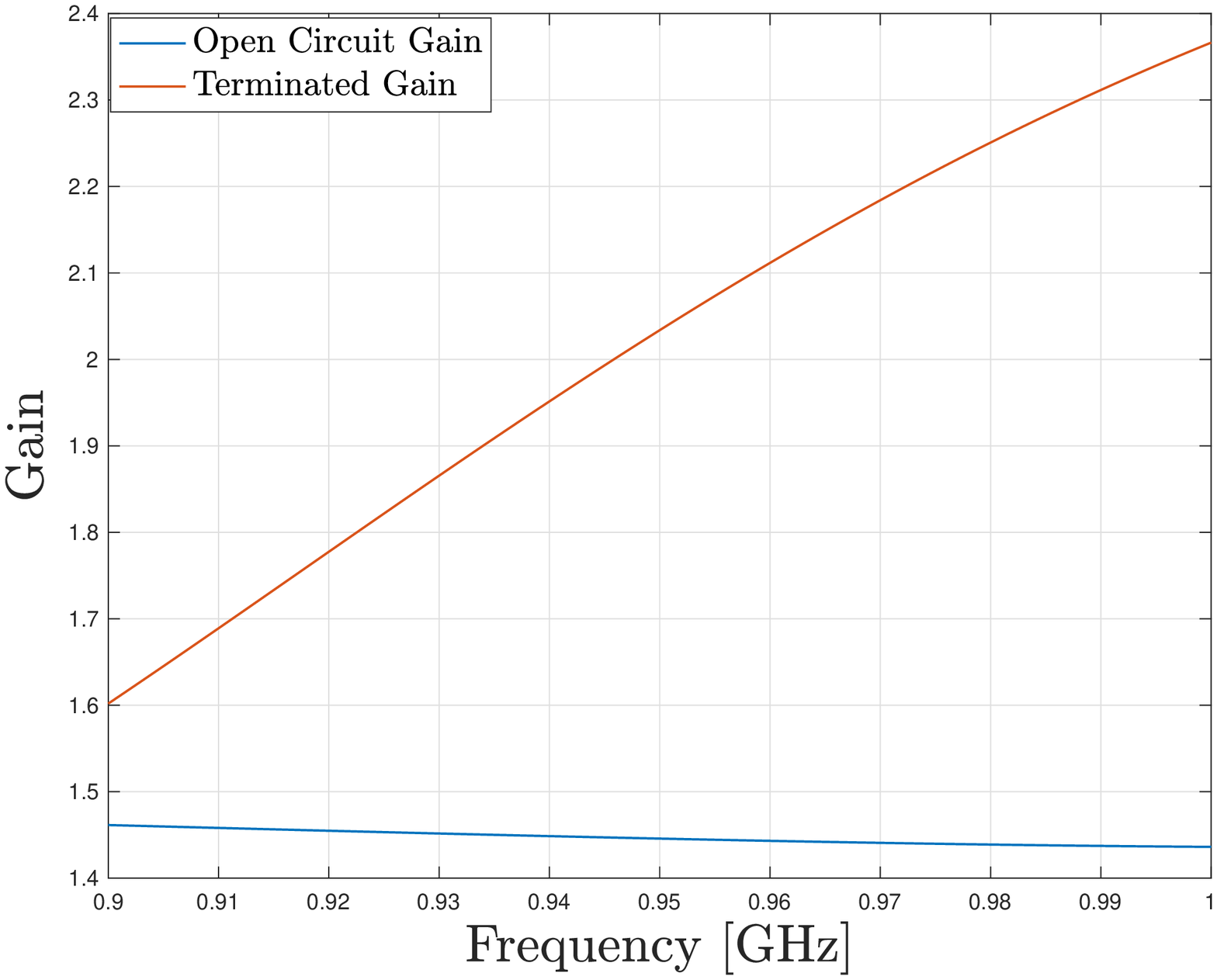}
  \caption{Open Circuit/Terminated Gain}
  \label{fig:Gain dipole}
\end{subfigure}%
\begin{subfigure}{.5\textwidth}
  \centering
  \includegraphics[width=1\linewidth]{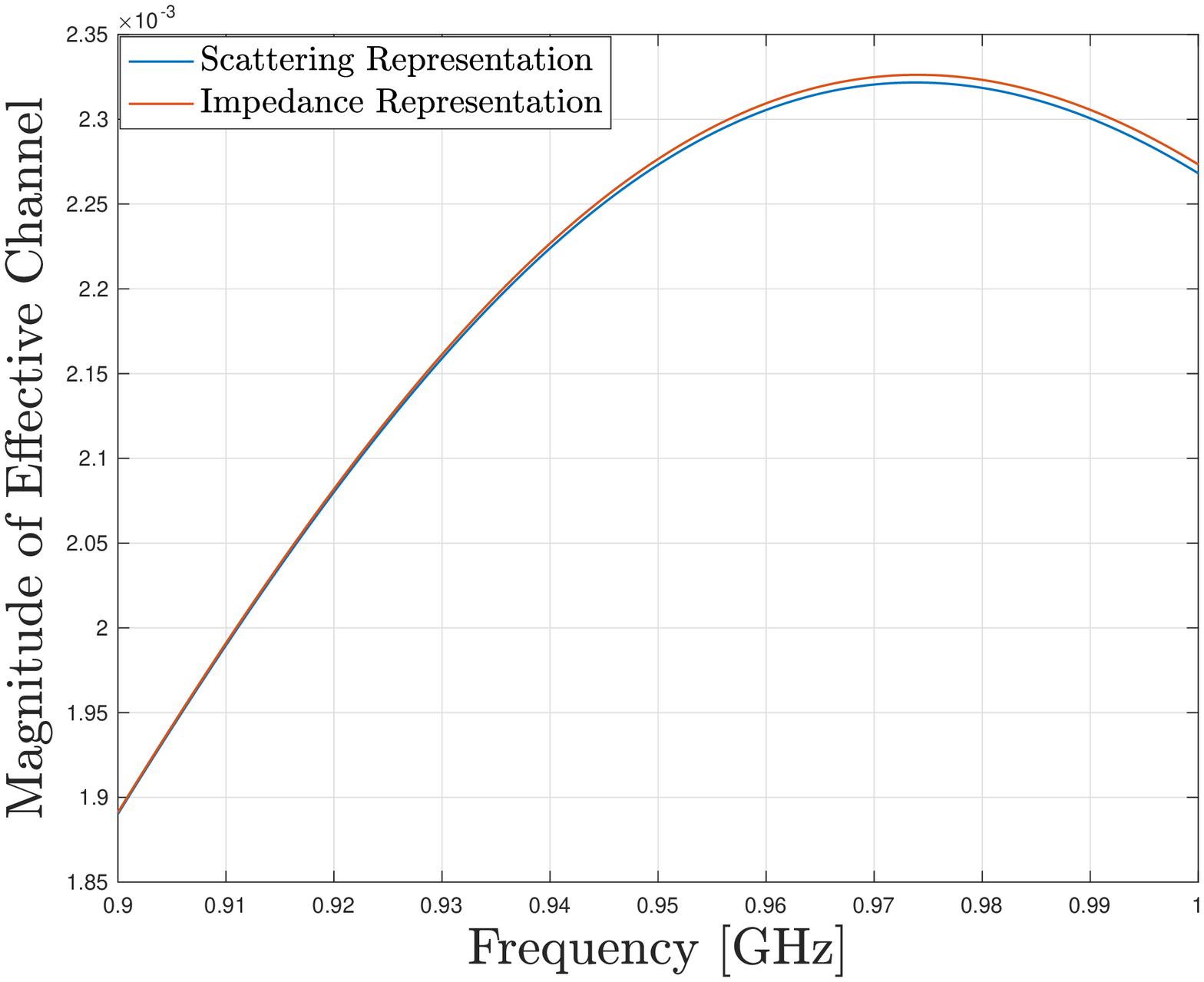}
  \caption{Effective channel magnitude from both representations}
  \label{fig:Channel power}
\end{subfigure}
\caption{Open circuit/Terminated antenna gains as well as effective channels}
\label{fig:Channel/Gain compare}
\end{figure}
where $G_{\rm OC/Term}$ stands for either an open circuit or a terminated gain, and $\mathbfcal{H}_{\rm OC/Term}(f)$ stands for the propagation channel from the impedance/scattering description. As can be seen in Fig.\ref{fig:Channel/Gain compare}, the terminated and open circuit embedded gains are very different, yet the description in (\ref{eq:Scattering_Channel_MIMO}) captures this difference theoretically (since $\mathbfcal{H}_{\rm eff}(f)$ is the same) effectively converting terminated gain into an open circuit gain.
\subsection{Performance Comparison}
\begin{figure}[h!]
    \centering
    \includegraphics[width=0.7\linewidth]{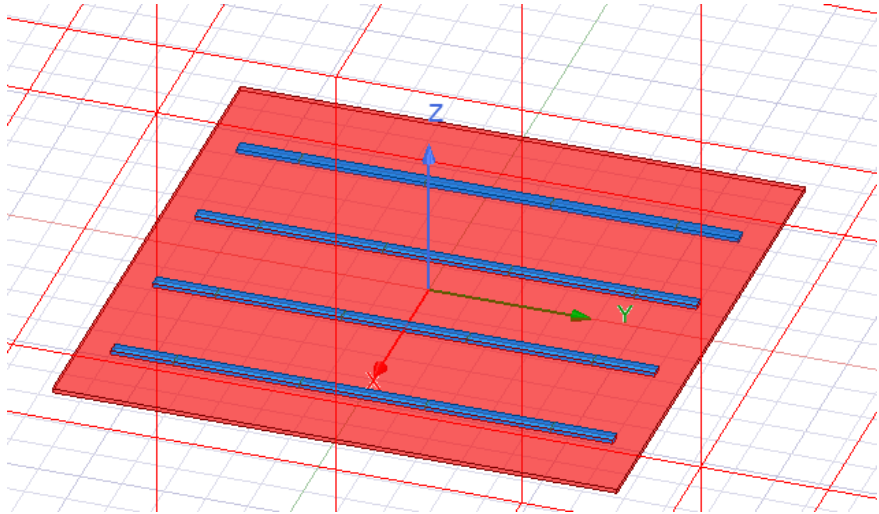}
    \vskip 0.1cm
    \caption{Connected array of slot antennas with 16 antenna elements} 
    \label{fig: Connected array}
\end{figure}

In this section, we compare the performance of the AB and AA LMMSE estimators. We restrict our attention to the connected array of slot antennas \cite{cavallo2011connected}, as such arrays are promising for next-generation wireless systems due to their wide bandwidth, large scan angles, and scalability. An image of the designed antenna array using HFSS \cite{HFSS} is shown in Fig. \ref{fig: Connected array}. The frequency band under investigation is $[0.5,~5]$ {GHz} where the antenna ports are separated by half-wavelength $\lambda_h/2$ at $5$ [GHz], the highest frequency of operation. The width of the slots is chosen to be $w = \lambda_h/60$. With $16$ antenna elements, the total length of the slot is $2\lambda_h$ where the metallic plate is taken as a square perfect electric conductor (PEC) of dimensions $(2\lambda_h + \lambda_h/4)\times(2\lambda_h + \lambda_h/4)$. The thickness of the plate is $t = w$. We also utilize a design with the connected array backed by a metallic plate of the same dimensions as the original plate placed at $\lambda_h/2$ below the antenna. 

\begin{figure}[h!]
    \centering
    \includegraphics[width=0.95\linewidth]{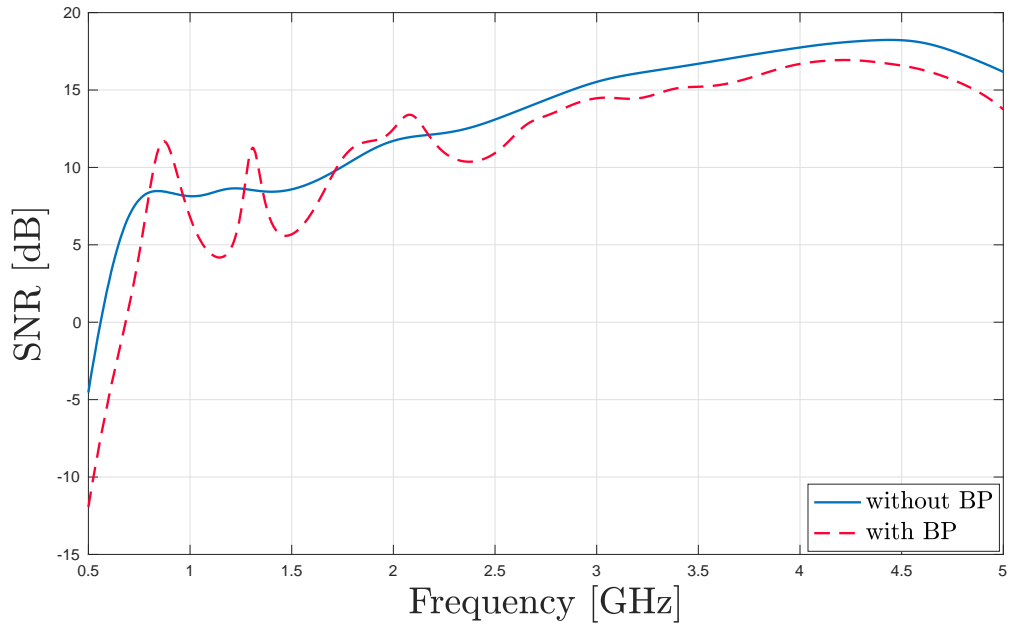}
    \caption{Empirical SNR against carrier frequency for the array with and without the backed-plane. Total Symbol Power is 1 [W] with a bandwidth of 5 [MHz] for each frequency point.}  
    \label{fig:SNR_vs_freq}
\end{figure}

\noindent For the large-scale parameter we use the extended Friis model:
\begin{equation}\label{eq:large_scale_parameter}
    \rho = \left(\frac{c}{4 \pi f d_{\mathrm{ref}}}\right)^{2}\left(\frac{d_{\mathrm{ref}}}{d}\right)^{\alpha},
\end{equation}
where $d_{\mathrm{ref}}$ is the reference pathloss distance, chosen as $1$ [m], and $\alpha$ is the pathloss exponent. We use i.i.d BPSK pilot signals in all simulations and unless stated otherwise we take $d$ = $100$ [m] and use a pathloss exponent of $\alpha$ = $2$ in all simulations. Defining the empirical single-carrier signal-to-noise ratio (SNR) as:
\begin{equation}\label{eq:SNR_definition}
    \textrm{SNR} ~\triangleq~ \frac{\mathbb{E}\{\| \sqrt{\rho}\,\mathbf{H}_{\textit{eff}}\mathbf{x}\|^{2}\}}{\mathbb{E}\{\| \mathbf{n}\|^{2}\}},
\end{equation}
we demonstrate the broadband properties of the arrays in use in Fig. \ref{fig:SNR_vs_freq} where we plot the SNR against carrier frequency with and without the backed-plane (BP) using 1000 Monte-Carlo (MC) simulations. We use a bandwidth of $5$ [MHz] with a symbol power of $1$ [W] at each frequency. 
Indeed we see that these arrays have an SNR above $5$ [dB] almost over the entire frequency range of interest. This is unlike narrowband antennas which will be resonant over a small band of frequencies. At low frequencies, we see a large degradation of SNR due to an inability to collect power because of the array's small physical area. We also note the enhanced frequency selectivity of the array with the backed plane due to the constructive/destructive reflection.
\newline 
\noindent We assess the performance of the two estimators using the normalized mean-square error (NMSE) as a performance metric:
\begin{equation}\label{eq:NMSE_definition}
    \textrm{NMSE}(\mathbf{\widehat{x}}) ~=~ \frac{\mathbb{E}{\{\|\mathbf{x} - \mathbf{\widehat{x}}\|^2}\}}{\mathbb{E}{\{\|\mathbf{x}\|^2}\}},
\end{equation}
where $\mathbf{x}$ is the theoretical value and $\mathbf{\widehat{x}}$ is the estimated value. We also measure the gains in spectral efficiency when using the AA estimator over the AB one by lower bounding the achievable rate.
\newline
\subsubsection{\textbf{Single-Carrier Transmission}}
\begin{figure}[h!]
    \centering
    \includegraphics[width=0.9\linewidth]{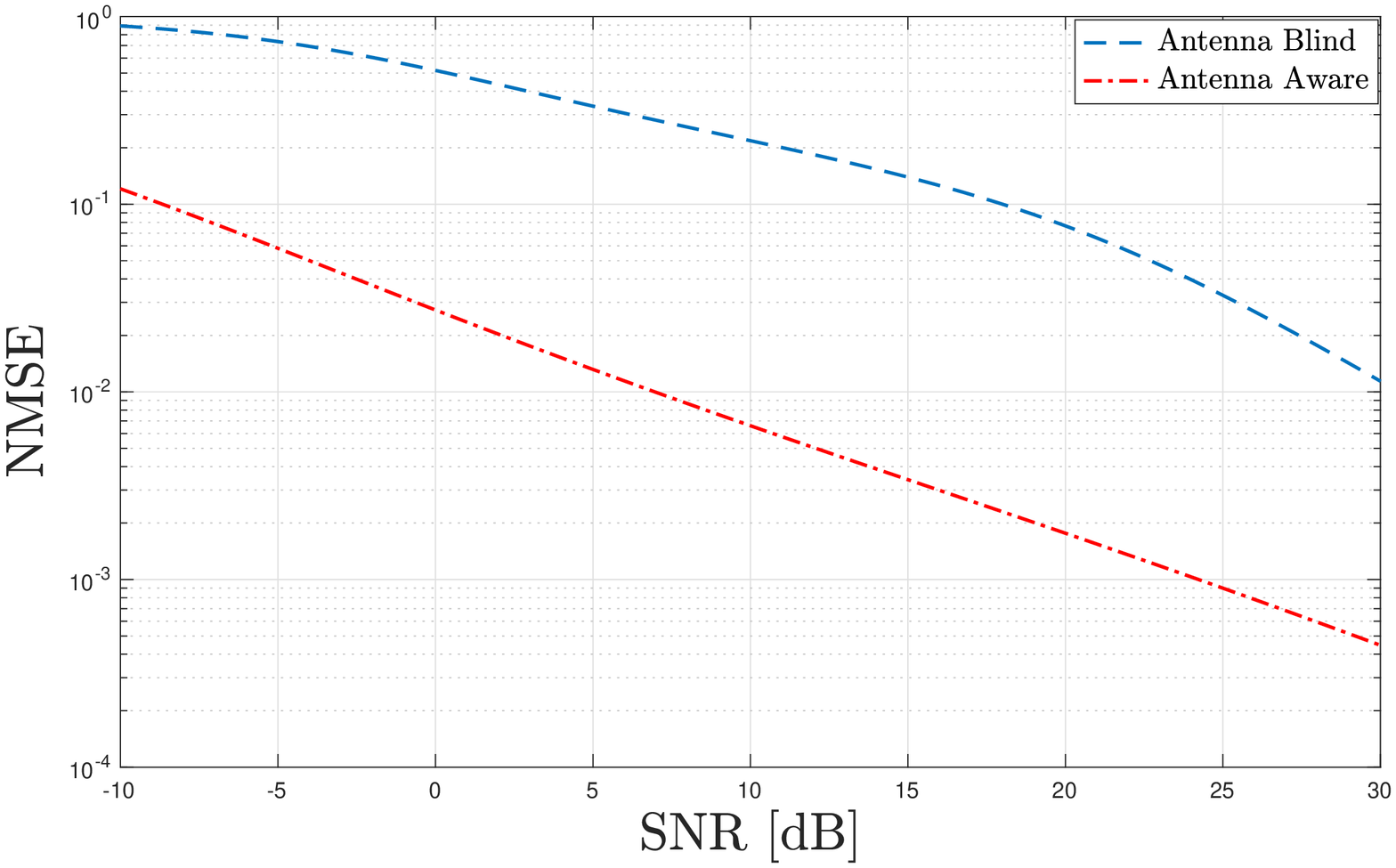}
    \caption{NMSE against pilot power at carrier frequency of 1 [GHz] with a bandwidth of $5$ [MHz].}  
    \label{fig:NMSE_vs_power}
\end{figure}

\begin{figure}
\centering
\begin{subfigure}{.5\textwidth}
  \centering
  \includegraphics[width=1\linewidth]{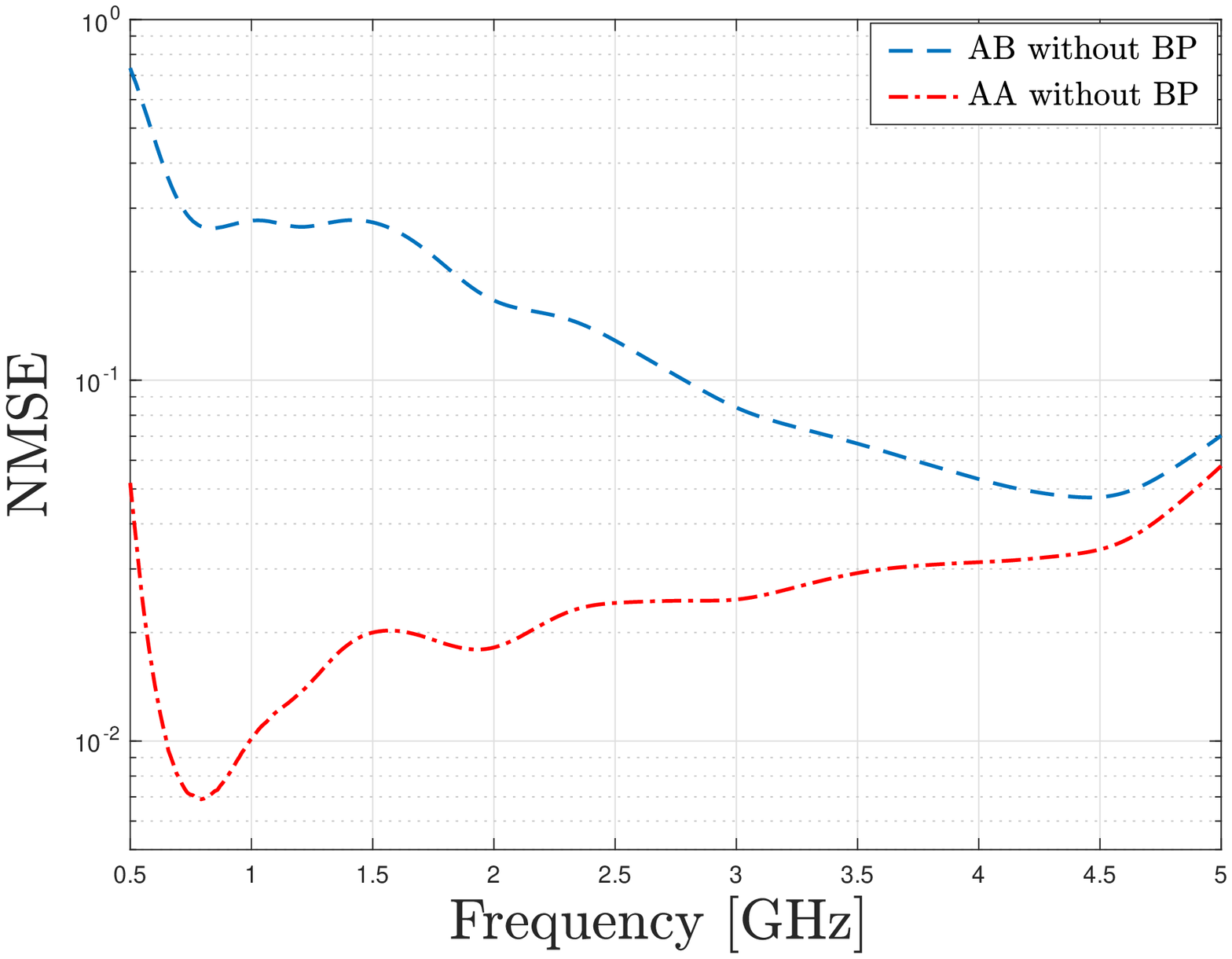}
  \caption{Array without the BP}
  \label{fig:NMSE_vs_freq_no_BP}
\end{subfigure}%
\begin{subfigure}{.5\textwidth}
  \centering
  \includegraphics[width=1\linewidth]{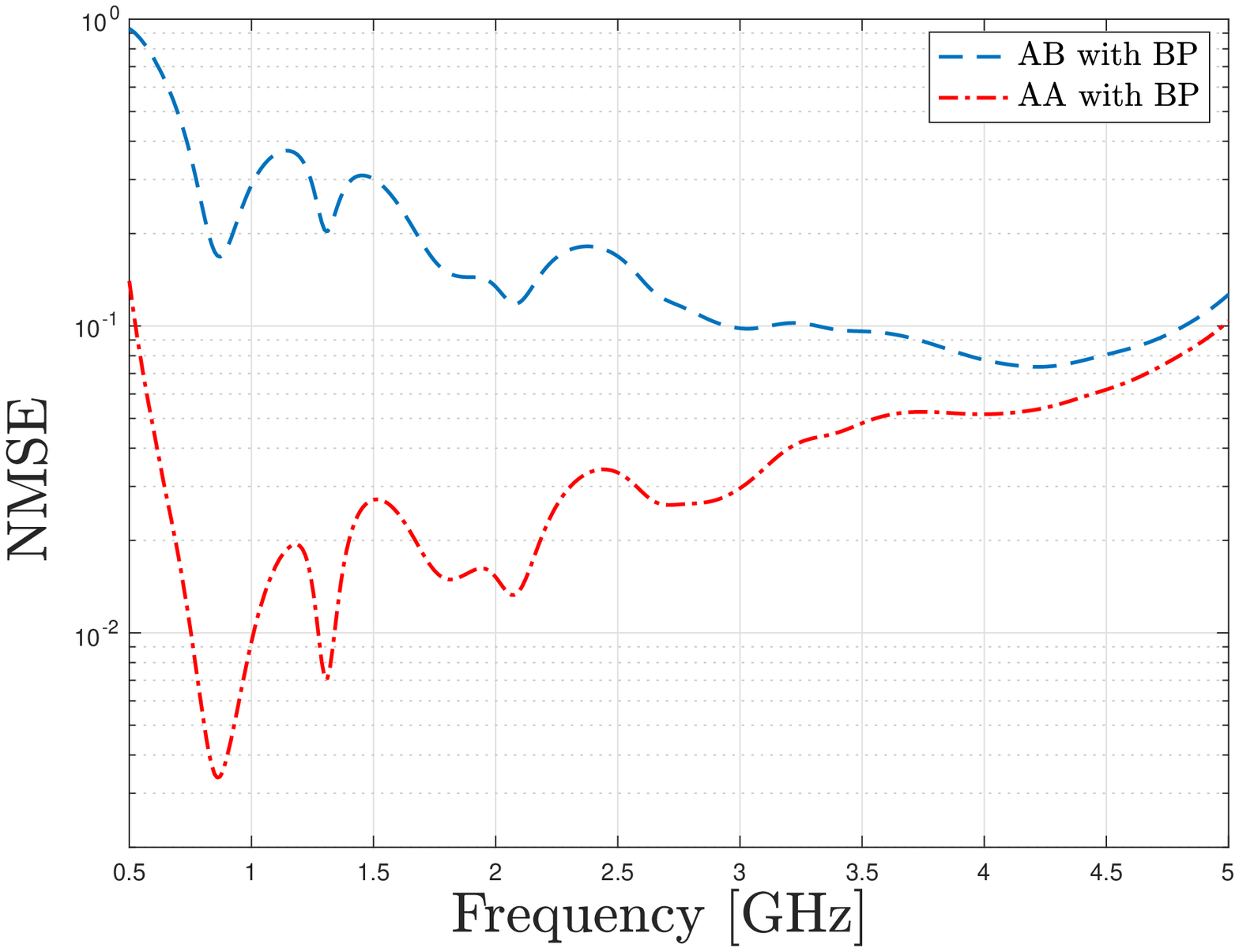}
  \caption{Array with BP}
  \label{fig:NMSE_vs_freq_BP}
\end{subfigure}
\caption{NMSE against carrier frequency where total power used in estimation for each frequency is 1 [W] with a bandwidth of 5 [MHz].}
\label{fig:NMSE_vs_Freq}
\end{figure}

\begin{figure}
\centering
\begin{minipage}{.5\textwidth}
  \centering
  \includegraphics[width=1\linewidth]{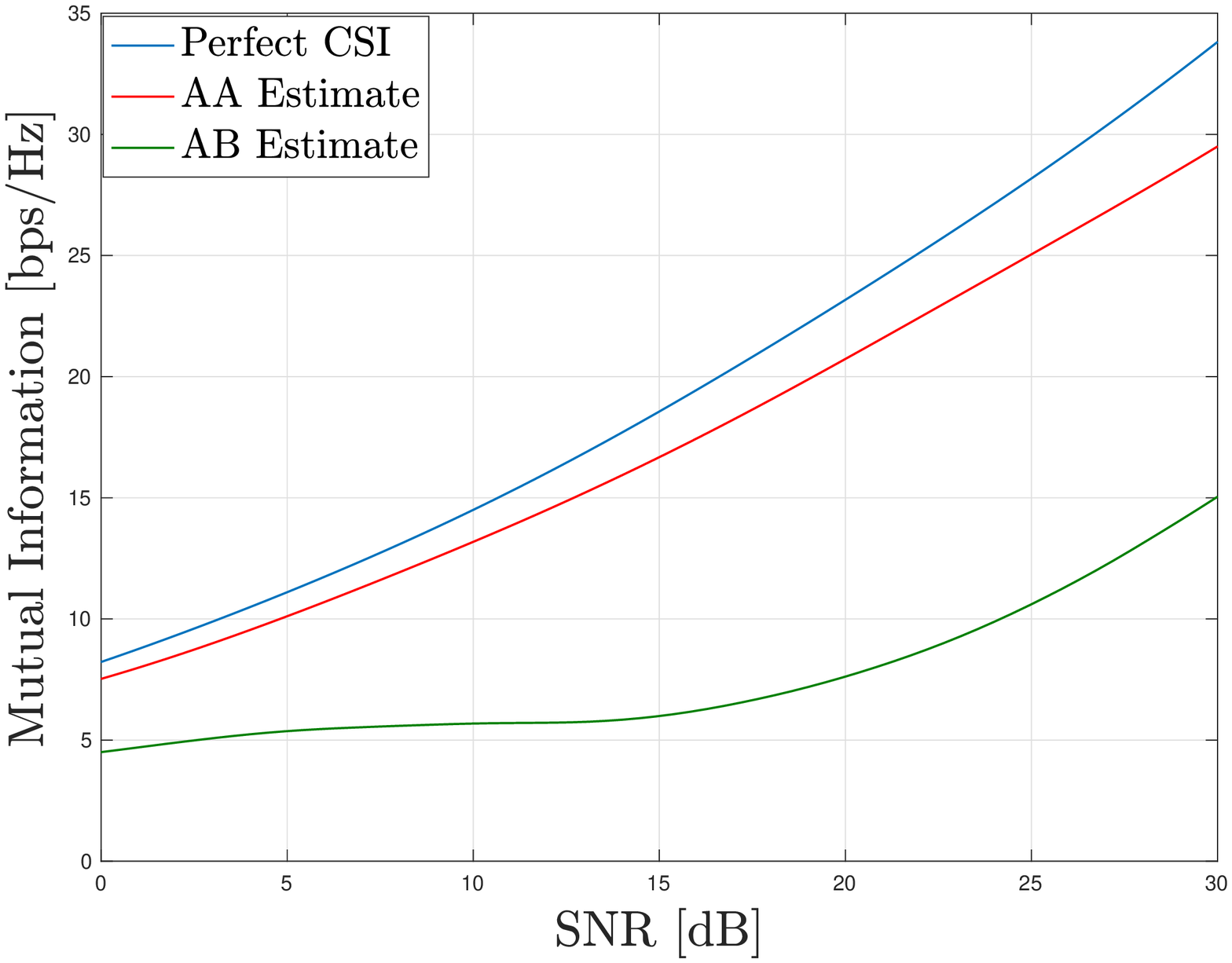}
  \captionof{figure}{Achievable rate at $f_c = 1$ [GHz].}
  \label{fig:Rate_vs_SNR_without_BP}
\end{minipage}%
\begin{minipage}{.5\textwidth}
  \centering
  \includegraphics[width=1\linewidth]{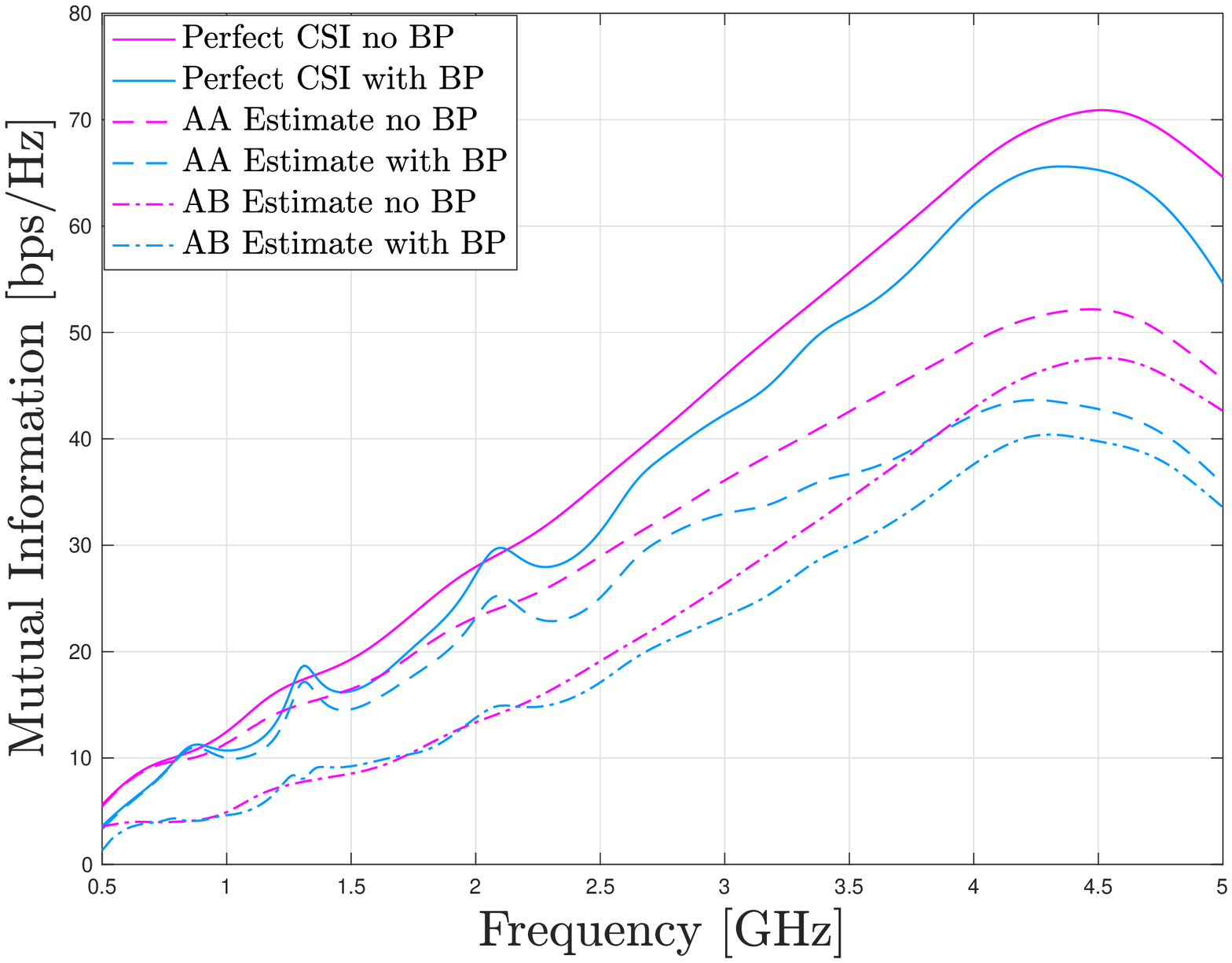}
  \captionof{figure}{Achievable rate against carrier frequency.}
  \label{fig:Rate_vs_frequency}
\end{minipage}
\end{figure}

We can compute the theoretical NMSE for the AB estimators using the MSE matrix (\ref{eq:SC_MSE_Matrix_effective}) and the covariance of the channel (\ref{eq:Kronecker_correlation_effective}):
\begin{equation}
    \textrm{NMSE}_{\textrm{AB}} ~=~ \frac{\mathrm{tr}(\mathbf{E}_{\textit{eff}}^{\textrm{AB}})}{\mathrm{tr}(\mathbf{R}_{\mathbf{H}_{\textit{eff}}})}.
\end{equation}
$\textrm{NMSE}_{\textrm{AA}}$ for the AA estimator is derived similarly using (\ref{eq:SC_MSE_Heff}) and (\ref{eq:Kronecker_correlation_effective}).
\noindent We plot the NMSE against pilot power in Fig. \ref{fig:NMSE_vs_power} for the AB and AA estimators using the connected array without the backed plane (similar results follow for the array with the BP). 
In this plot we run $1000$ MC to compute the SNR and use $L_t = 20$ time-slots to estimate the channel. This is done at a carrier frequency of $1$ [GHz] with a bandwidth of $5$ [MHz]. We see more than 20 [dB] improvement in the NMSE for the AA estimator as compared to the AB estimator. At high SNR we see the gains start to decrease as the increase in pilot power overcomes the mismatched covariance assumptions. We next plot the NMSE against carrier frequency for the array with and without the BP in Fig. \ref{fig:NMSE_vs_Freq}.
In these plots, we use a bandwidth of $5$ [MHz] and total pilot power of $1$ [W]. We see that as the carrier frequency increases the gain of the AA estimator decreases. This is because the fixed spacing of the array elements, $\lambda_h / 2$, starts to become a significant fraction of the carrier wavelength, $\lambda_c$, as the carrier frequency increases, and thus the coupling between antennas becomes negligible. We also see that our AA estimator brings in most gains in the frequency range of $0.6$-$1$ [GHz] where the array is well matched and the mutual coupling is significant.
\\
\indent Next, we study the impact of channel estimation on the achievable rate (a key figure-of-merit in communication systems). To do this, we denote the estimated effective channel using the AB and AA estimators by $\mathbf{\widehat{H}}_{\textit{eff}}^{\mathrm{AB}}$ and $\mathbf{\widehat{H}}_{\textit{eff}}^{\mathrm{AA}}$ respectively. We assume that the channel estimate is available at both the transmitter and receiver, this can occur in a time-division-duplex (TDD) system with a reciprocal channel. As our goal is solely to investigate the impact of channel estimation we assume that after estimation both the AA and AB methods are aware of the full model in (\ref{eq: Discrete time model}). Without loss of generality, we first whiten the noise by multiplying (\ref{eq: Discrete time model}) by $\mathbf{L}^{-1}$ (recall $\mathbf{R}_{\mathbf{n}} = \mathbf{L}\mathbf{L}^{\mathsf{H}}$). For the purposes of precoding and processing at the receiver we define the singular-value decomposition (SVD) of the product of $\mathbf{L}^{-1}$ and the estimated channels by:
\begin{equation}\label{eq:SVD_estimates}
\begin{aligned}
    \mathbf{L}^{-1}\mathbf{\widehat{H}}_{\textit{eff}}^{\mathrm{AB}}&~\triangleq~\mathbf{U}^{\mathrm{AB}}\mathbf{\Sigma}^{\mathrm{AB}}(\mathbf{V}^{\mathrm{AB}})^{\mathsf{H}},\\
    \mathbf{L}^{-1}\mathbf{\widehat{H}}_{\textit{eff}}^{\mathrm{AA}}&~\triangleq~\mathbf{U}^{\mathrm{AA}}\mathbf{\Sigma}^{\mathrm{AA}}(\mathbf{V}^{\mathrm{AA}})^{\mathsf{H}}.
\end{aligned}
\end{equation}
For the AB system\footnote{The same process is done for the AA system} we use $\mathbf{V}^{\mathrm{AB}}$ to precode the transmit signal according to:
\begin{equation}
    \mathbf{x} = \sqrt{\frac{P_{T}}{N_t}}\mathbf{V}^{\mathrm{AB}}\begin{bmatrix}
    (\mathbf{P}_{\mathrm{AB}})^{1/2}\\
    \mathbf{0}
    \end{bmatrix}
    \mathbf{s},
\end{equation}
in which $P_T$ is the transmit power and $\mathbf{s} \in \mathbb{C}^{N_s\times 1}$ is the information bearing signal whose components contain $N_s$ independently coded data streams. The matrix $(\mathbf{P}_{\mathrm{AB}})^{1/2}$ is a diagonal matrix containing the square root of the set of optimal powers allocated across the $N_s$ data streams based on the singular values from $\mathbf{\Sigma}^{\mathrm{AB}}$. These are found using the water-filling policy under a per-symbol average power constraint, i.e., $\mathrm{tr}(\mathbf{P}) = N_t$. In addition to multiplying the received signal by $\mathbf{L}^{-1}$ we also pre-process it by multiplying by $(\mathbf{U}^{\mathrm{AB}})^{\mathsf{H}}$. If the estimates $\mathbf{U}^{\mathrm{AB}}$ and $\mathbf{V}^{\mathrm{AB}}$ were perfect, this process would yield a bank of parallel subchannels with a sum capacity given by:
\begin{equation}
    C(\mathbf{H}_{\textit{eff}}) ~=~ \sum_{j=0}^{N_{\textrm{min}}-1}\log_{2}(1+\mathrm{SNR}_{j}),
\end{equation}
where $\mathrm{SNR}_{j} = \frac{\rho P_{T}\sigma_{j}^2 P_{j}^{*}}{N_t}$, $\sigma_{j}$ is the $j^{th}$ singular value of $\mathbf{L}^{-1}\mathbf{H}_{\textit{eff}}$, and $P_{j}^{*}$ is the power allocated to the $j^{th}$ stream.  Channel estimation errors, however, introduce inter-stream interference which is independent of the background noise. By treating such interference as one more additive noise term while still decoding the streams separately, the 
mutual information between $\mathbf{s}$ and $\mathbf{y}$ (scaled accordingly to get units of bits/s/Hz) of this system given $\mathbf{H}_{\textit{eff}}$ is lower bounded by \cite{heath2018foundations}:
\begin{equation}
    I(\mathbf{\widehat{H}}_{\textit{eff}}^{\mathrm{AB}} | \mathbf{H}_{\textit{eff}}) ~\triangleq~ \sum_{j=0}^{N_{\textrm{min}}-1}\log_{2}(1+\mathrm{SINR}_{j}^{\textrm{AB}}),
\end{equation}
where 
\begin{equation}\label{eq:SINR_single_carrier}
 \mathrm{SINR_{j}^{\textrm{AB}}} = \frac{\frac{\rho P_{T} }{N_t} P_{j}^{\textrm{AB}^{*}} |(\mathbf{u}_{j}^{\mathrm{AB}})^{\mathsf{H}}(\mathbf{L}^{-1}\mathbf{H}_{\textit{eff}})\mathbf{v}_{j}^{\mathrm{AB}} |^{2} }{1 ~+~ \frac{\rho P_{T} }{N_t} \sum_{l \neq j} P_{l}^{\textrm{AB}^{*}} |(\mathbf{u}_{j}^{\mathrm{AB}})^{\mathsf{H}}(\mathbf{L}^{-1}\mathbf{H}_{\textit{eff}})\mathbf{v}_{l}^{\mathrm{AB}} |^{2}},
\end{equation}
and $P_{l}^{\textrm{AB}^{*}}$ is the power allocated to the $l^{th}$ stream using the waterfilling algorithm based on the AB singular values in (\ref{eq:SVD_estimates}). In Fig. \ref{fig:Rate_vs_SNR_without_BP}  we plot the average achievable rate against SNR without the backed-plane (results are similar with the BP). This was done at $f_c= 1$ [GHz], a bandwidth of $5$ [MHz] and with $1000$ MC simulations. 
At moderate to high SNR, when the noise effects are not dominant we see a gain of about $10-15$ [bpcu].
Next in Fig. \ref{fig:Rate_vs_frequency} we plot the achievable rate against the carrier frequency with and without the backed plate. 
Again, we see the that the array with the backed plane is more frequency selective and has performance degradation compared to the one without the backed plate. The largest gains for both arrays are found around $2$ [GHz] where there is significant mutual coupling. At larger frequencies we do not see much gains using the antenna aware estimator as the mutual coupling becomes insignificant. We also see that the gap to the perfect CSI case increases. This is due to the fact that the SNR (see Fig. \ref{fig:SNR_vs_freq}) becomes large at higher frequencies and the decrease in interference in the denominator of (\ref{eq:SINR_single_carrier}) is negligible compared to the increase in SNR given by the numerator of (\ref{eq:SINR_single_carrier}).
\newline
\subsubsection{\textbf{Multicarrier Transmission}}
\begin{figure}[h!]
    \centering
    \includegraphics[width=1\linewidth]{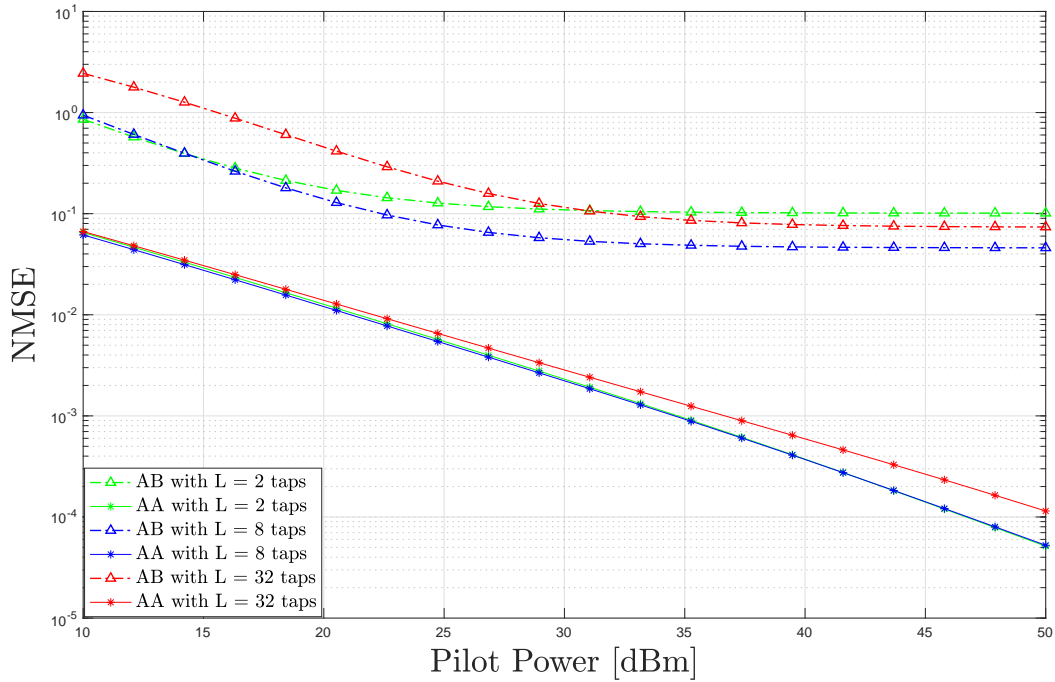}
    \caption{NMSE against pilot power in OFDM Estimation using total bandwidth of 708 [MHz]}  
    \label{fig:OFDM_NMSE_vs_Pilot_power}
\end{figure}

\begin{figure}[ht]
    \centering
    \includegraphics[width=0.95\linewidth]{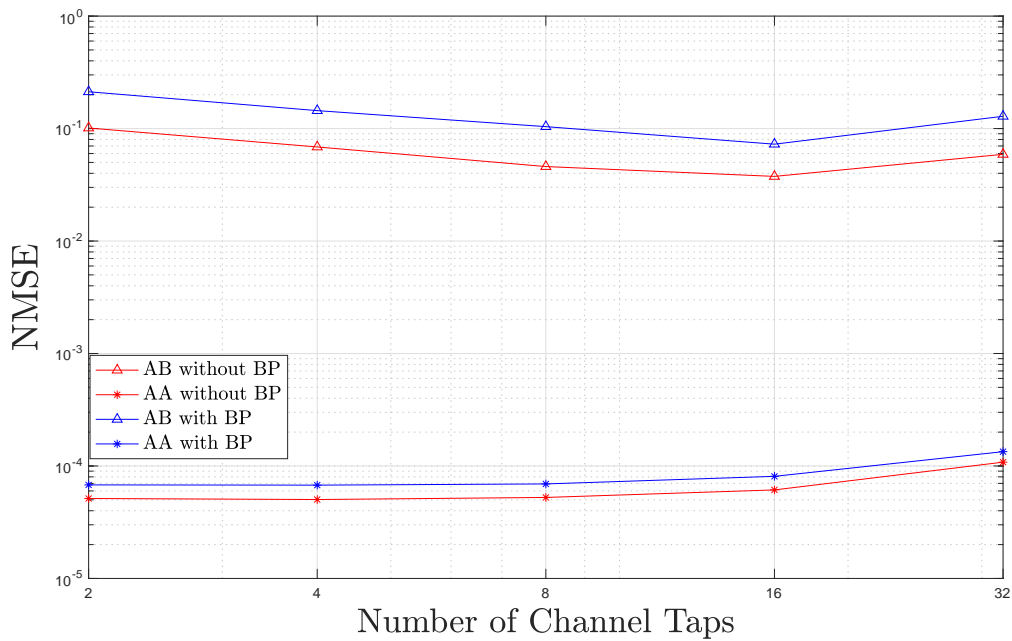}
    \caption{NMSE against number of taps at high SNR with and without the BP}  
    \label{fig:OFDM_NMSE_vs_taps}
\end{figure}

\begin{figure}
\centering
\begin{subfigure}{.5\textwidth}
  \centering
  \includegraphics[width=1\linewidth]{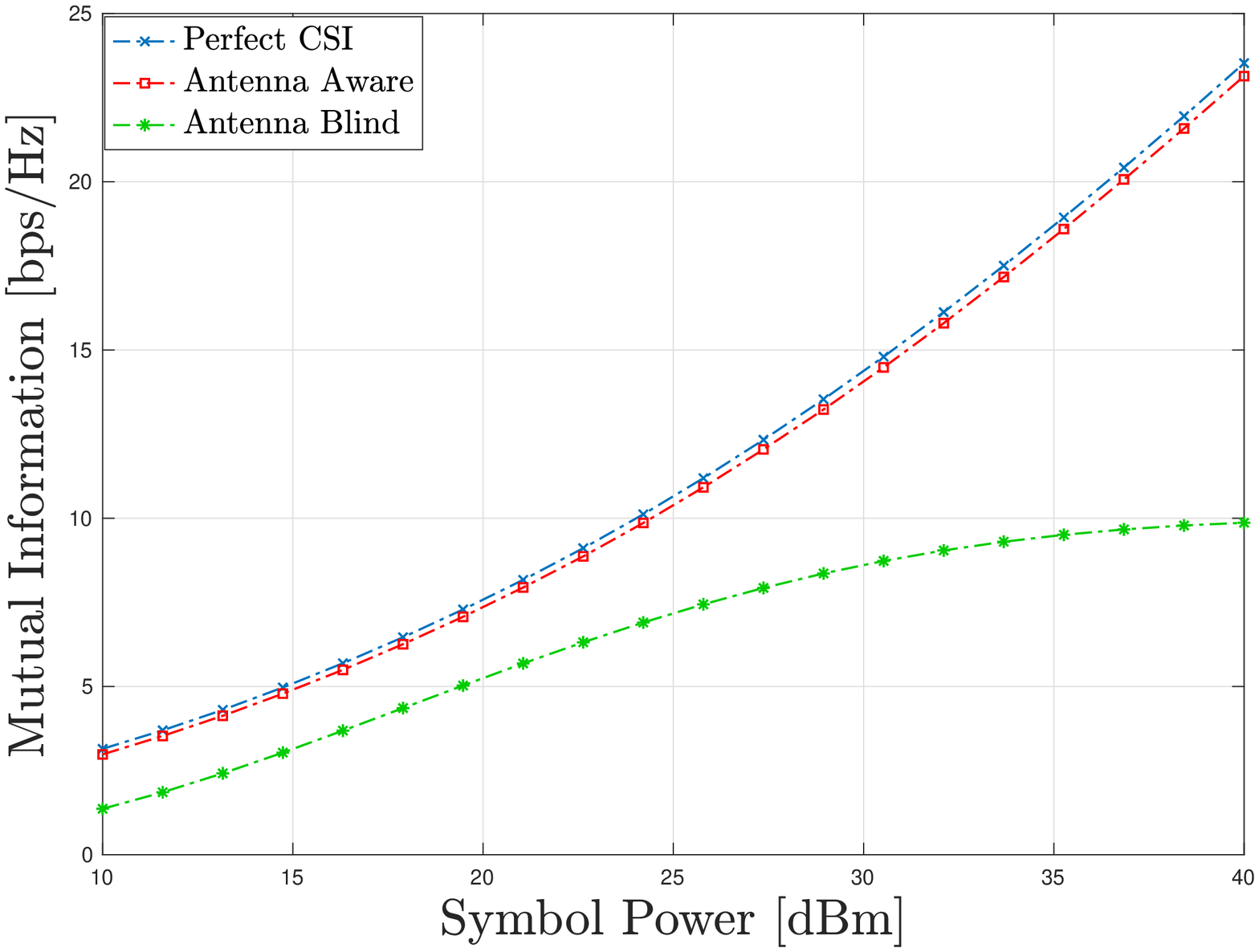}
  \caption{Achievable Rate against Power}
  \label{fig:OFDM_rate_vs_power}
\end{subfigure}%
\begin{subfigure}{.5\textwidth}
  \centering
  \includegraphics[width=1\linewidth]{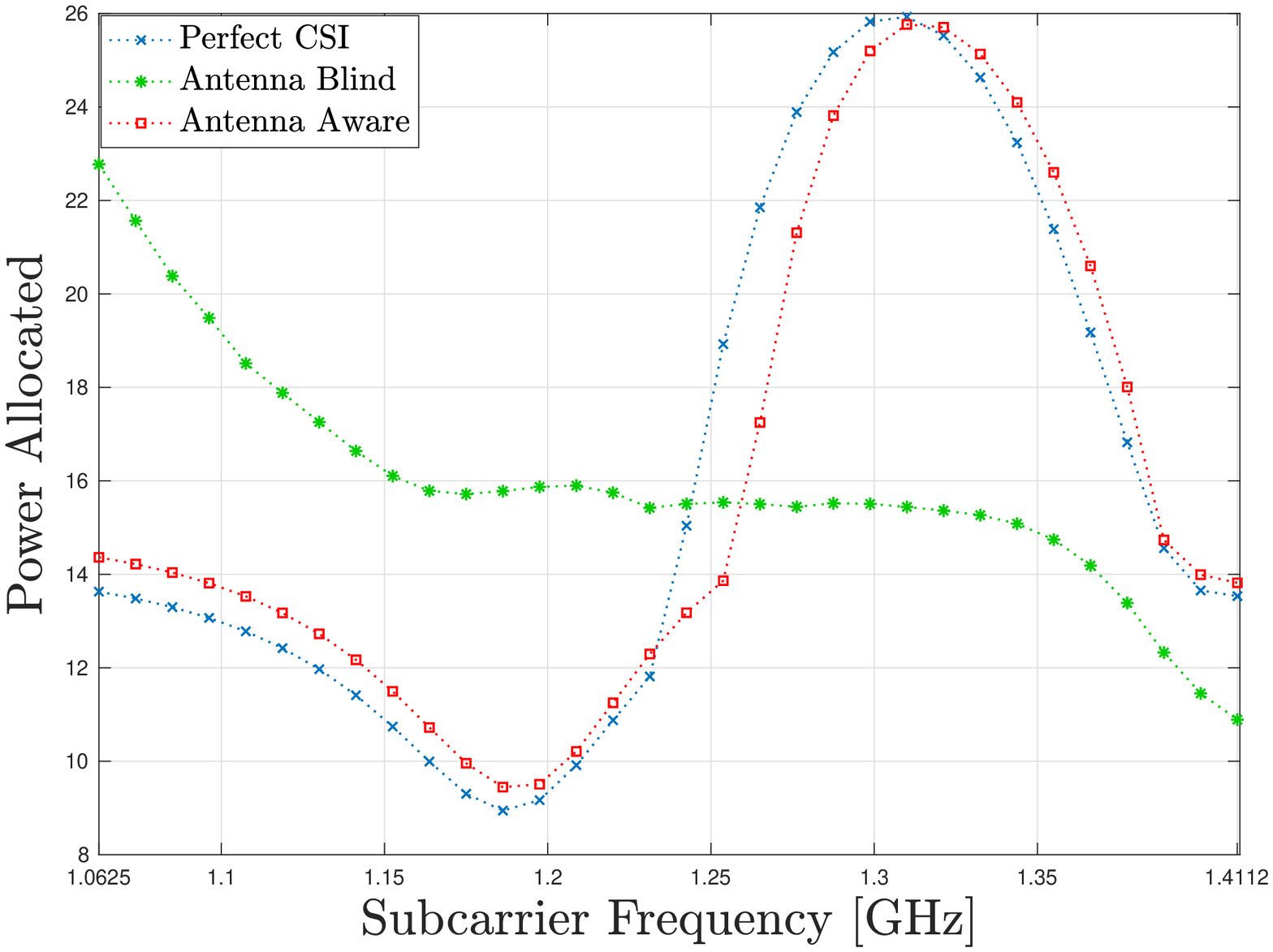}
  \caption{Power Allocation over Subcarriers}
  \label{fig:OFDM_PA_vs_freq}
\end{subfigure}
\caption{Achievable rate plot and power allocation plot using the backed-plane array with $64$ subcarriers over the frequency band $1-1.5$ [GHz] with $2$ channel taps.}
\label{fig:NMSE_vs_Freq_2}
\end{figure}

The OFDM AB (\ref{eq:OFDM_AB_estimator}) and AA (\ref{eq:OFDM_AA_estimator}) estimators will provide an estimate of the given $L$-tap MIMO channels in time given by (\ref{eq:OFDM_AB_estimate}) and (\ref{eq:OFDM_AA_estimate}) respectively. The theoretical NMSE for the AB and AA estimators can be found using (\ref{eq:OFDM_AB_MSE}), (\ref{eq:OFDM_MSE_AA_eff}), and (\ref{eq:Covariance_channel_freq}):
\begin{equation}\label{eq:OFDM_NMSE_defn}
\begin{aligned}
    \textrm{NMSE}_{\textrm{AB}} &=~ \frac{\mathrm{tr}(\mathbf{E}_{\textit{eff}}^{\textrm{AB}})}{\mathrm{tr}(\mathbf{R}_{\bm{\mathsf{H}}_{\textit{eff}}})},\\
    \textrm{NMSE}_{\textrm{AA}} &=~ \frac{\mathrm{tr}(\mathbf{E}_{\textit{eff}}^{\textrm{AA}})}{\mathrm{tr}(\mathbf{R}_{\bm{\mathsf{H}}_{\textit{eff}}})}.
\end{aligned}
\end{equation}
The large scale parameter (\ref{eq:large_scale_parameter}) depends on the frequency of operation and in broadband OFDM systems this can vary significantly with the subcarrier frequency. As this can be estimated easily at the receiver we add the frequency dependence in the pilot symbols so that we can use one common large-scale parameter as in (\ref{eq:OFDM_AB_estimator}) and (\ref{eq:OFDM_AA_estimator}). In Fig. \ref{fig:OFDM_NMSE_vs_Pilot_power} we plot the the NMSE against the pilot power parameterized by the number of channel taps using 10 monte-carlo for each data point and $10$ time-domain pilots. This is done over the frequency band from [$1$ GHz, $1.8$ GHz] with a total bandwidth of 800 [MHz] with 64 subcarriers.
At low SNR, for the AB estimator, we see that the NMSE increases as the number of taps increase. Note here at $10$ [dBm] of power the NMSE for the $8$-tap channel is higher than the $2$-tap channel, and decreasing the power more we will see a bigger gap between the two. This low SNR behaviour is due to the fact that the model mismatch effects become negligible as the noise is dominant. Therefore, for a given power, it is easier to estimate a smaller number of variables. At high SNR however, the model mismatch effects become dominant. Indeed, we see that the $2$ tap channel has the highest NMSE followed by the $32$ and $8$ taps channel. Again, if we increase the pilot power more we will see that the $32$-tap channel will have the lowest NMSE. Here this is due to the fact that increasing the assumed number of taps more accurately captures the frequency selectivity introduced by the antennas, i.e., the model is more correctly matched. For the AA estimator, we see that the NMSE for $2$ and $8$ tap channels are almost the same over all SNRs and that the $32$ tap channel is also the same except at high SNR, where the matrix conditioning in the estimator  (\ref{eq:OFDM_AA_estimator}) gets worse with more taps leading to noise enhancement effects. This is further demonstrated in Fig. \ref{fig:OFDM_NMSE_vs_taps} where we plot the NMSE against the number of channel taps ($L$). This is done at high SNR (pilot power $50$ [dBm]) where we can see the full effects of the model mismatch. As confirmed in Fig. \ref{fig:SNR_vs_freq} the array with the backed plane is more frequency selective and hence should introduce more taps than the array in free space. Due to this, in Fig. \ref{fig:OFDM_NMSE_vs_taps}, we see the gains using the AA estimator over the AB estimator are greater for the backed array with the highest gains at a low number of taps consistent with intuition.
\newline
Similar to the single-carrier case we can assess the AA and AB estimators using the achievable rate performance criteria. Due to the properties of OFDM transmission, a frequency selective channel can be converted to a bank of $K$ parallel subchannels given by:
\begin{equation}\label{eq:OFDM_achievable_rate_model}
    \bm{\mathsf{y}}[k] ~=~ \sqrt{\rho}\bm{\mathsf{H}}_{\textit{eff}}[k]\,\bm{\mathsf{x}}[k] + \bm{\mathsf{n}}[k].
\end{equation}
One approach to obtain the achievable rate is to allocate power uniformly over the subcarriers and then apply the waterfilling algorithm over space. However, when using broadband arrays this approach will be sub-optimal due to the enhanced frequency selectivity of the arrays. Therefore, to optimize the mutual information, we jointly allocate power over space and frequency. This is first done by whitening the noise for each subcarrier in (\ref{eq:OFDM_achievable_rate_model}) and taking the SVD, i.e.,
\begin{equation}   
    \bm{\mathsf{L}}^{-1}[k]\bm{\mathsf{{H}}}_{\textit{eff}}[k]~\triangleq~\bm{\mathsf{U}}[k]\,\bm{\mathsf{\Sigma}}[k]\,\bm{\mathsf{V}}^{\mathsf{H}}[k].
\end{equation}
Next the product $\bm{\mathsf{L}}^{-1}[k]\bm{\mathsf{{H}}}_{\textit{eff}}[k]$ can be concatenated in a block-diagonal matrix over the $K$ subcarriers. The SVD of this newly constructed block diagonal matrix is a block-diagonal matrix of the $\bm{\mathsf{U}}[k]$'s multiplied by a block-diagonal matrix of the $\bm{\mathsf{\Sigma}}[k]$'s multiplied by a block-diagonal matrix of the $\bm{\mathsf{V}}[k]$'s. Now based on this block-diagonal matrix of the $\bm{\mathsf{\Sigma}}[k]$'s we can perform power allocation under a per-symbol power constraint over all subcarriers equal to $K\,N_t$. We precode based off the blocks of the block diagonal $\bm{\mathsf{V}}[k]$'s and process the received signal by blocks of the block diagonal $\bm{\mathsf{U}}[k]$'s. If this process was done based on imperfect CSI then we can lower bound the mutual information by treating the interference as Gaussian noise. 
To get the achievable rate in bpcu we must sum over the number of subcarriers and normalize by the number of subcarriers as they are transmitted serially. The lower bound of the mutual information for the AB estimator, in bpcu, is given by:
\begin{equation}
    I(\{\mathbf{\widehat{H}}_{\textit{eff}}^{\mathrm{AB}}\}_{k=0}^{K-1} | \{\mathbf{H}_{\textit{eff}}\}_{k=0}^{K-1}) ~\triangleq~ \frac{1}{K}\sum_{k=0}^{K-1}\sum_{j=0}^{N_{\textrm{min}}-1}\log_{2}(1+\mathrm{SINR}_{kj}^{\textrm{AB}}),
\end{equation}
where 
\begin{equation}
 \mathrm{SINR_{kj}^{\textrm{AB}}} = \frac{\frac{\rho_{k} P_{T} }{N_t} P_{kj}^{\textrm{AB}} |(\mathbf{u}_{kj}^{\mathrm{AB}})^{\mathsf{H}}(\mathbf{L}^{-1}\mathbf{H}_{\textit{eff}})\mathbf{v}_{kj}^{\mathrm{AB}} |^{2} }{1 ~+~ \frac{\rho_{k} P_{T} }{N_t} \sum_{l \neq j} P_{kl}^{\textrm{AB}} |(\mathbf{u}_{kj}^{\mathrm{AB}})^{\mathsf{H}}(\mathbf{L}^{-1}\mathbf{H}_{\textit{eff}})\mathbf{v}_{kl}^{\mathrm{AB}} |^{2}}.
\end{equation}
In Fig. \ref{fig:OFDM_rate_vs_power} we plot the achievable rate against power using the backed-plane array with $64$ subcarriers from $1-1.5$ [GHz] and $2$ channel taps.
We see that achievable rate obtained using the AA channel estimate is very close to perfect CSI as the estimation error is very small. We also see that at higher SNR the gap between the AB achievable rate and the perfect CSI grow. This is due to the model mismatch effects that become prominent at higher SNR as explained previously. In Fig. \ref{fig:OFDM_PA_vs_freq} we plot the total power allocated for each subcarrier against the subcarrier frequency using a symbol power of $10$ [dBm]. Indeed, the power's allocated for the perfect CSI case and the AA case follow the SNR plot of the backed-plane array over the band $1-1.5$ [GHz] seen in Fig. \ref{fig:SNR_vs_freq} demonstrating the utility in our approach. Furthermore, we see the effect that an inaccurate channel estimate can have as the powers allocated for the AB estimator have no resemblance to the SNR in Fig. \ref{fig:SNR_vs_freq}.

\section{Conclusion}
 In this paper, we developed a novel LMMSE channel estimator for single- and multi-carrier systems that takes advantage of the mutual coupling in the transmit/receice antennae arrays. We model the mutual coupling through multiport networks and express the single-user  MIMO communication channel in terms of the impedance and scattering parameters of the antenna arrays. In frequency-flat single-carrier systems, we show that neglecting the coupling in the arrays leads to  an inaccurate characterization of  the channel and noise correlations. In frequency-selective multi-carrier channels, we show this same effect and also demonstrate that the coupling in the arrays will increase the number of channel taps. Standard LMMSE estimators developed under these inaccurate models become sub-optimal and hence we develop an LMMSE estimator that calibrates the coupling and optimally estimates the channel. It is shown that appropriately accounting for mutual coupling through the developed physically consistent model leads to remarkable improvements in terms of channel estimation performance. We demonstrate the gains in our algorithm in a rich-scattering environment using a connected array of slot antennas both at the transmitter and receiver sides. In the future research, it would be useful to explore the design of the pilot sequences to match the antenna rather than choosing orthogonal pilots as is done conventionally. Other avenues could be to explore the channel-sparsity as well as the estimation of the AoA/AoD.



\bibliographystyle{IEEEtran}
\bibliography{IEEEabrv,references}
\end{document}